\documentclass[floatfix,reprint,amsmath,amssymb,aps,prb]{revtex4-2}
\usepackage{booktabs}
\usepackage{multirow}
\usepackage[american]{babel}
\usepackage{graphicx}
\usepackage{bm}
\usepackage[hidelinks]{hyperref}
\usepackage[nice]{nicefrac}
\usepackage{physics}
\usepackage{xcolor}
\usepackage{xspace}
\usepackage{mathtools} 
\usepackage[normalem]{ulem}
\usepackage{hyperref}
\usepackage{soul}
\usepackage{amsmath}
\usepackage{amssymb}
\usepackage{bm}
\usepackage[capitalize]{cleveref}

\hypersetup{
    colorlinks,
    bookmarksopen,
    bookmarksnumbered,
    citecolor=teal,
    linkcolor=teal,
    pdfstartview=Fit,
    urlcolor=teal
}

\begin{document}
\date{\today}

\title{Excitonic Stripe Order in the Two-Orbital Hubbard-Kanamori Model}

\author{Rafael D. Soares}
\thanks{These authors contributed equally to this work.}
\affiliation{Max Planck Institute for the Physics of Complex Systems, N\"{o}thnitzer Stra{\ss}e 38, 01187 Dresden, Germany}
\author{Luke Staszewski}
\thanks{These authors contributed equally to this work.}
\affiliation{Max Planck Institute for the Physics of Complex Systems, N\"{o}thnitzer Stra{\ss}e 38, 01187 Dresden, Germany}
\author{Chunhan Feng}
\affiliation{Max Planck Institute for the Physics of Complex Systems, N\"{o}thnitzer Stra{\ss}e 38, 01187 Dresden, Germany}
\author{Alexander Wietek}
\affiliation{Max Planck Institute for the Physics of Complex Systems, N\"{o}thnitzer Stra{\ss}e 38, 01187 Dresden, Germany}

\begin{abstract}
Excitonic condensation and stripe formation are two distinct manifestations of electronic correlations. While excitonic order naturally arises in multi-orbital systems, stripe order is a prominent feature of doped correlated-electron models. Here, we investigate an excitonic analogue of stripe order in the two--orbital Hubbard–Kanamori model on the square lattice, characterized by a spatial modulation of inter-orbital particle–hole coherence and, in the orbital-parity-symmetric limit, spontaneous breaking of a relative orbital \(\mathbb Z_2\) symmetry. Using unrestricted real-space Hartree–Fock calculations complemented by random-phase-approximation instability analysis, we determine how the Kanamori interactions select different excitonic channels. The sign of Hund’s exchange controls the spin character of the condensate: ferromagnetic Hund coupling favors triplet excitonic order, whereas antiferromagnetic Hund coupling stabilizes singlet excitonic order. Upon doping, commensurate excitonic density waves develop incommensurate textures, including excitonic stripes and, in the triplet sector, spiral excitonic density waves. We further show that crystal-field splitting strongly reorganizes the excitonic instability by modifying inter-orbital nesting and can stabilize triplet excitonic order even in the absence of pair hopping. Our results establish excitonic stripes as a distinct symmetry-broken state of multi-orbital correlated systems and identify microscopic routes for their stabilization.
\end{abstract}

\maketitle

\tableofcontents

\section{Introduction}
\label{sec:introduction}

A defining feature of strongly correlated electron systems is the profound competition between kinetic energy, which promotes itinerancy, and Coulomb repulsion, which drives localization and the formation of ordered phases. Mott insulators, unconventional superconductors and non-Fermi liquids are expected to be appropriately captured by a single-orbital Hubbard model~\cite{Hubbard1963,White1989,LeBlanc2015}. While this remains the standard approach for materials where a single orbital effective model reproduces many physical phenomena, like the cuprates~\cite{cui2025ab}, its conceptual simplicity fails to capture the complex ordered phases observed in materials that inherently rely on multi-orbital degrees of freedom. Indeed, multi-orbital physics is an essential consideration, as most known families of superconductors are fundamentally multi-orbital materials. This includes prominent correlated systems such as iron-based superconductors~\cite{yu2013orbital} and, notably, both infinite-layer (such as NdNiO$_2$)~\cite{li2019superconductivity} and Ruddlesden-Popper (such as La$_3$Ni$_2$O$_7$)~\cite{sun2023signatures} nickelates. Furthermore, this paradigm naturally extends to other layered compounds that have been intensely studied, such as transition metal dichalcogenides~\cite{Manzeli2017} and ruthenates~\cite{RevModPhys.75.657,kaba2019group}. In these systems, orbital degrees of freedom promote fundamentally different types of correlation~\cite{Medici2011_2,Fernandes2022}. This arises from the extra contribution of inter-orbital Coulomb interactions, which disfavor the simultaneous occupancy of multiple orbitals, and Hund’s coupling, which promotes the formation of correlated high-spin states~\cite{Georges2013,Capone2026}.
Rapid progress has recently established Hund’s coupling as the main mechanism behind the differentiation of correlation strengths among electrons belonging to different orbitals~\cite{Nicola2013,Medici2014}. This coupling suppresses orbital fluctuations~\cite{Medici2011,Medici2011_2}, which can lead to orbital locking~\cite{Kotliar2017} or, more remarkably, give rise to distinct forms of orbital-selective physics~\cite{Medici2005}. A paradigmatic example is the orbital-selective Mott phase, where a localized Mott insulating state in one orbital coexists alongside completely itinerant electrons in the others. However, more generically, orbital-resolved quasiparticles become fundamentally differentiated, exhibiting vastly distinct quasiparticle weights~\cite{Nicola2013,Kostin2018,Akiyama2026}. Another striking consequence of this physics is the enhancement of charge compressibility, which ultimately culminates in a Fermi-liquid instability toward phase separation~\cite{deMedici2017}.
Beyond the formation of Hund's metals, multi-orbital models also provide a rich framework for stabilizing excitonic phases of matter~\cite{Rademaker2013,Kune2015,Kaneko2026}. This stems from the interplay between Hund's coupling and inter-orbital density-density repulsion, which can drive the pairing of an electron in one orbital with a hole in another. When a macroscopic number of these particle-hole pairs acquire quantum phase coherence, they form an excitonic condensate, which is favored to be a spin-triplet state due to Hund's rules~\cite{Hund1925}. The resulting triplet excitonic (Et) condensate represents a macroscopic quantum state that spontaneously breaks both spin and orbital symmetries. Density functional theory calculations have predicted that Et phases exist in proximity of a spin-state transition in certain cobalt oxide materials, such as $\text{Pr}_{0.5}\text{Ca}_{0.5}\text{CoO}_3$~\cite{KUne2014,Yamaguchi2017} and $\text{La}\text{CoO}_3$~\cite{Sotnikov2017,wang2018}. Other noteworthy candidate materials are the $\text{Ta}_2 \text{NiSe}_5$~\cite{Wakisaka2009,Kim2021} and $1\text{T-TiSe}_2$~\cite{Cercellier2007} compounds, even though the exact nature of the instability towards the excitonic phase is still under debate, as it is suspected that it might not be driven by strong electronic interactions alone, but rather by coupling to phononic modes or due to structural transitions~\cite{Hedayat2019,Mazza2020,Pashov2025}. In two-orbital models with broken orbital degeneracy, recent sign-free quantum Monte Carlo studies have shown that an antiferromagnetic Hund's coupling stabilizes singlet excitonic (Es) order~\cite{Huang2022}, while a ferromagnetic coupling drives Et order. Moreover, recent theoretical work suggests that this exotic phase can be realized dynamically as a non-equilibrium steady state through photo-doping of a multi-orbital Mott insulator~\cite{Geng2026,Yan2026}.

\begin{figure}[t]
    \centering
    \includegraphics[width=\linewidth]{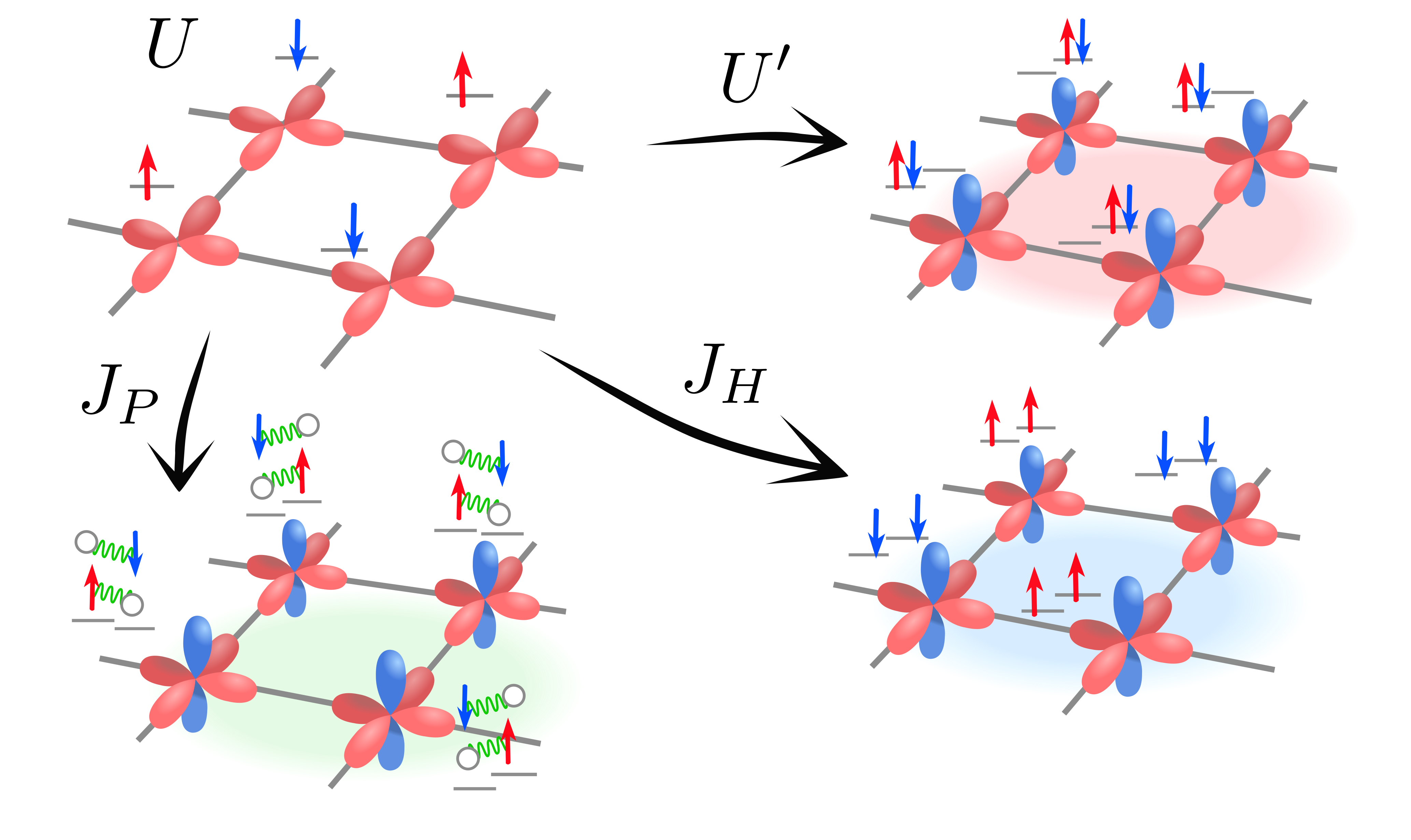}
    \caption{
    Schematic illustration of the principal ordered phases that emerge in the generalization of the single orbital Hubbard model to the two-orbital Hubbard-Kanamori model at the Hartree–Fock level: Néel order (blue), orbital density wave order (red); and excitonic order (green). }
    
    \label{fig:schematic}
\end{figure}

These multi-orbital effects are naturally described within the Hubbard-Kanamori model~\cite{Kanamori1963}. Although this model has been extensively investigated, previous studies have largely focused on either ground-state properties or on the physically relevant parameter regime dictated by the rotational symmetries of the partially filled atomic $d$-shells, where the interaction parameters are not independent. Here, we adopt a different perspective. Rather than targeting a specific material, we systematically investigate the two--orbital Hubbard-Kanamori model on a square lattice at finite temperature, treating the interaction parameters as independent variables. While the Mermin-Wagner theorem~\cite{PhysRevLett.17.1133} precludes true long-range continuous symmetry breaking in 2D at finite temperatures, the Hartree-Fock analysis typically captures the dominant ordering tendencies and the onset of short-range correlations near the zero temperature limit. This approach allows us to disentangle the roles of the intra- and inter-orbital Coulomb interactions, Hund's exchange, and pair hopping in stabilizing the competing ordered phases of the model as schematically shown in ~\cref{fig:schematic}. Taking the spherically symmetric limit of the local Coulomb interactions~\cite{197038,Georges2013} as our starting point, a regime whose magnetic properties closely resemble those of the single-orbital Hubbard model, we continuously explore the extended interaction space and uncover a zoo of orders, including several excitonic phases. Our results show how the different components of the Kanamori interaction compete and cooperate to stabilize distinct symmetry-broken ordered phases of multi-orbital correlated systems.

We begin in ~\cref{sec:hamiltonian_methods} by introducing the two--orbital Hubbard-Kanamori model and discussing its internal symmetries. By constructing the complete set of local, linearly independent Hermitian one-body operators, we classify the possible symmetry-broken phases and their order parameters according to the irreducible representations of the full symmetry group (\cref{sec:orders_multi_orbital}). Guided by random-phase approximation (RPA) calculations in the high-temperature phase, we determine which instability dominates near the critical mean-field temperature for different model parameters and map out temperature--electron-density phase diagrams for representative parameter sets in \cref{sec:temp_doping_phase_diagrams}. We then systematically explore the broader parameter space in \cref{sec:parameter_scans}, where transitions between competing ordered phases emerge beyond the instability of the high-temperature state. Finally, \cref{sec:crystal_field_and_t_inter} demonstrates how crystal-field splitting and inter-orbital hopping reduce the orbital symmetry, which constrains the possible symmetry-broken phases but allows for excitonic condensates at material realistic parameters. We summarize our main conclusions in \cref{sec:conclusions}, with further methodological details deferred to the Appendices.

\begin{figure*}[t]
    \centering
    \includegraphics[width=\linewidth]{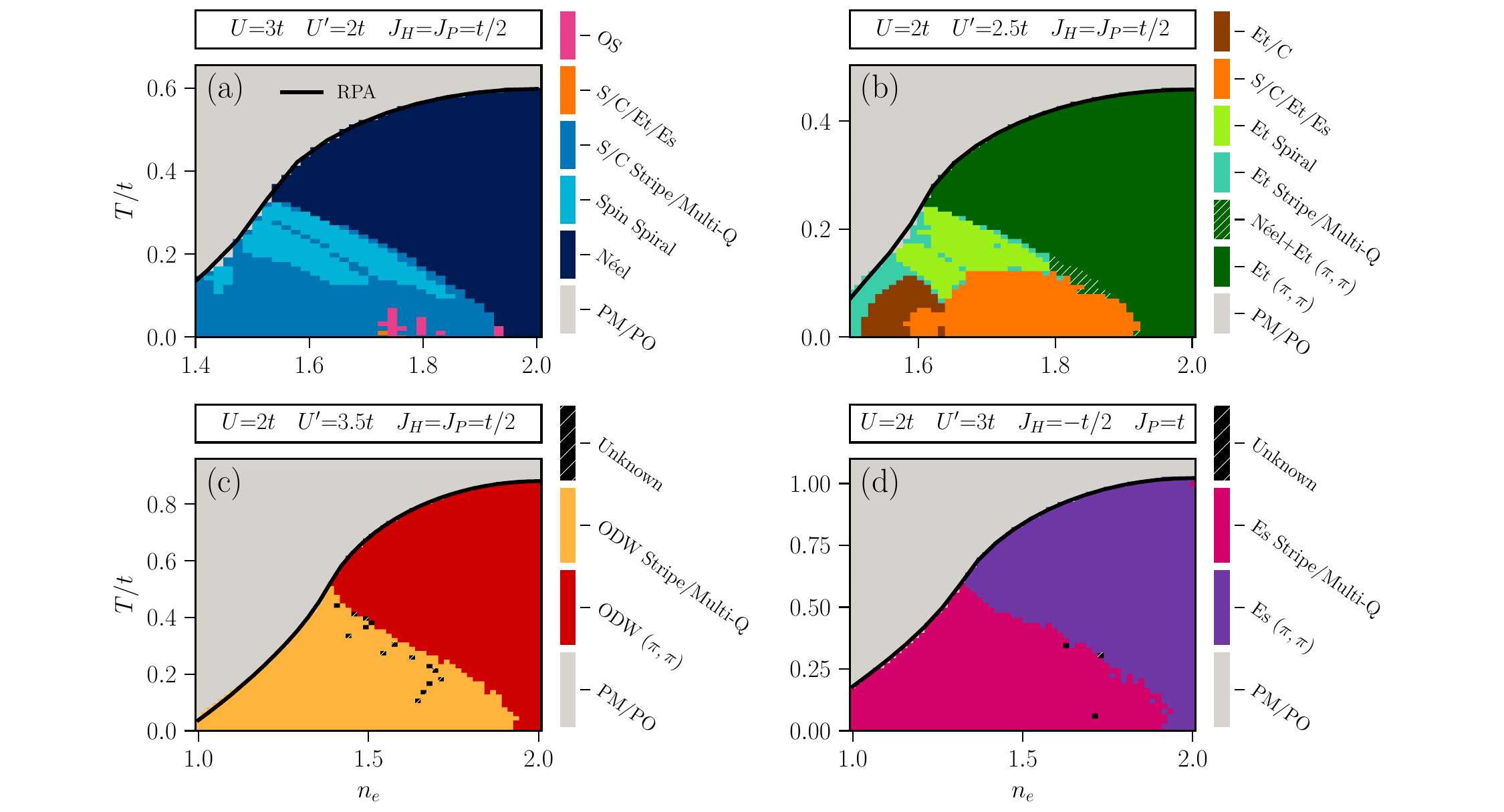}
    \caption{Temperature-electron density phase diagrams for four characteristic regimes of the Hubbard-Kanamori model. Panel shows (a) the predominantly intra-orbital magnetically ordered regime ($U=3.0t$, $U^\prime=U-2J_H$, $J_H=J_P=0.5t$); (b) $E_t$ ordered phase ($U=2.0t$, $U^\prime=2.5t$, $J_H=J_P=0.5t$); (c) orbitally ordered region ($U=2.0t$, $U^\prime=3.5t$, $J_H=J_P=0.5t$); and (d) $E_s$ ordered regime ($U=2.0t$, $U^\prime=3.0t$, $J_H=-0.5t$, $J_P=t$). Black lines indicate phase boundaries derived from spin and charge susceptibility instabilities in the paramagnetic (PM) and para-orbital (PO) states within the RPA. Labels S and C denote spin and charge, respectively, with the $/$ symbol indicating potential coexistence. Fixed parameters across all panels are $L=20$ and $\Delta_{\rm cf}=t_{\rm inter}=0$.}
    \label{fig:T_vs_n_magnetic_dome_w_JP}
\end{figure*}

\section{The Hubbard-Kanamori Hamiltonian} 
\label{sec:hamiltonian_methods}

We consider the two--orbital Hubbard-Kanamori model on a two-dimensional square lattice with $N = L^2$ sites. The full Hamiltonian is defined as
\begin{equation}
\mathcal{H} = \mathcal{H}_0 + \mathcal{V}.
\end{equation}
The non-interacting part of the Hamiltonian, $\mathcal{H}_0$, consists of intra-orbital nearest-neighbor hopping with amplitude $t$, on-site inter-orbital hybridization with strength $t_{\rm inter}$ and an orbital-dependent on-site energy, the crystal-field splitting $\Delta_{\rm cf}$, which originates from the local electrostatic potential of the lattice and lifts the orbital degeneracy. The Hamiltonian reads,
\begin{equation}
\begin{aligned}
\mathcal{H}_0 &= \sum_{j,\alpha,\sigma} \left(\varepsilon_{\alpha} - \mu \right)\;n_{j,\alpha,\sigma}  - t_{\rm inter} \sum_{j,\alpha\neq\gamma,\sigma} c_{j,\alpha,\sigma}^{\dagger} c_{j,\gamma,\sigma}\\
&-t\sum_{\langle i,j \rangle,\alpha,\sigma} \left(c^\dagger_{i,\alpha,\sigma} c_{j,\alpha,\sigma} +\text{h.c.}\right),
\end{aligned}
\end{equation}
where $\mu$ corresponds to the chemical potential, $c_{j,\alpha,\sigma}^{\dagger}$ ($c_{j,\alpha,\sigma}$) is the creation (annihilation) operator of a fermion on the lattice site $j$ with spin $\sigma \in \{\uparrow, \downarrow\}$ and orbital index $\alpha \in \{1, 2\}$, and $n_{j,\alpha,\sigma} = c_{j,\alpha,\sigma}^{\dagger} c_{j,\alpha,\sigma}$ is the number operator. The crystal-field splitting is parametrized by the on-site energies, $\varepsilon_1 = -\Delta_{\rm cf}/2$ and $\varepsilon_2=\Delta_{\rm cf}/2$. 

The interacting part of the Hamiltonian $\mathcal{V}$ is given by the Kanamori interaction~\cite{Kanamori1963},
\begin{equation}
\begin{aligned}
\mathcal{V} &= U \sum_{j,\alpha} n_{j,\alpha,\uparrow} n_{j,\alpha,\downarrow} + \left(U^\prime -\frac{J_H}{2} \right) \sum_{j,\sigma,\sigma^\prime} n_{j,1,\sigma}n_{j,2,\sigma^\prime} \\
&- 2J_H\sum_{j} \boldsymbol{S}_{j,1} \cdot \boldsymbol{S}_{j,2} + J_P \sum_{j} \left(P^\dagger_{j,1} P_{j,2} + \text{h.c.} \right),
\end{aligned}
\label{eq:kanamori_interaction}
\end{equation}
where $\boldsymbol{S}_{j,\alpha}=\frac{1}{2} \sum_{\mu,\nu}c^\dagger_{j,\alpha,\mu}\boldsymbol{\sigma}_{\mu\nu}c_{j,\alpha, \nu}$ is the spin-$1/2$ operator for orbital $\alpha$, and $P_{j,\alpha}=c_{j,\alpha,\downarrow}c_{j,\alpha,\uparrow}$ is the intra-orbital pair operator. The interaction parameters include an on-site intra-orbital Hubbard repulsion of strength $U$, an inter-orbital density-density repulsion $U^\prime-J_H/2$, an on-site inter-orbital Heisenberg exchange with coupling $J_H$, and an inter-orbital pair-hopping term with amplitude $J_P$.

We now present a systematic analysis of the internal symmetries of the two-orbital Hubbard-Kanamori Hamiltonian. To the best of our knowledge, a generic discussion of the full symmetry group, including orbital-parity and exchange symmetry, has not been coherently discussed in the literature. Establishing the symmetry group is necessary to identify the order parameters that characterize distinct physical phases, which we detail in the \cref{sec:orders_multi_orbital}.

For a generic value of the parameters in the Kanamori interaction, the Hamiltonian always has a global $\text{U(1)}$ symmetry associated with the conservation of the total particle number, $N=\sum_{j,\alpha,\sigma} n_{j,\alpha,\sigma}$, and a $\text{SU(2)}$ spin-rotation symmetry since the orbital independent transformation above leaves the Hamiltonian invariant, 
\begin{equation}
    \mathcal{U}\, c_{j,\alpha,\sigma}\, \mathcal{U}^{-1} = e^{-i\theta} \sum_{\sigma^\prime} \mathbb{U}_{\sigma \sigma^\prime} c_{j,\alpha,\sigma^\prime},
\end{equation}
where $\mathbb{U}\in \text{SU}(2)\subseteq \mathbb{C}^{2\times2}$ and $\theta\in \mathbb{R}$.

On the other hand, the symmetry group of the orbital degrees of freedom, $G_o$, strongly depends on the specific value of the parameters. For $t_{\rm inter} = J_P = 0$, $G_o=\text{U}(1)$ as the difference between the particle number within each orbital is individually conserved. If $t_{\rm inter} = 0$, while the pair-hopping amplitude $J_P$ remains finite, this continuous symmetry reduces to a discrete $\mathbb{Z}_2$ symmetry, since only the relative parity of the occupation number per orbital remains a good quantum number~\cite{Huang2022}. The action of this symmetry is implemented by the transformation,
\begin{equation}
    \mathcal{P} c_{j,1,\sigma} \mathcal{P}^{-1} = c_{j,1,\sigma}, \quad \mathcal{P} c_{j,2,\sigma} \mathcal{P}^{-1} = -c_{j,2,\sigma}.
    \label{eq:relative_parity}
\end{equation}
A finite inter-orbital hopping, $t_{\rm inter}$, explicitly breaks this symmetry. Consequently, in the fully generic case ($t_{\rm inter}, \Delta_{\rm cf}, J_P \neq 0$), the orbital sector lacks any continuous or discrete symmetries. Setting the crystal-field splitting to zero ($\Delta_{\rm cf}=0$) yields another $\mathbb{Z}_2$ symmetry, as the model is invariant under the permutation of the two orbitals,
\begin{equation}
    \mathcal{X} c_{j,1,\sigma} \mathcal{X}^{-1} = c_{j,2,\sigma}, \quad \mathcal{X} c_{j,2,\sigma} \mathcal{X}^{-1} = c_{j,1,\sigma}.
\end{equation}
As such, when $t_{\rm inter} =\Delta_{\rm cf} =0$ but $J_P \neq 0$, the full orbital symmetry group forms the dihedral group of order eight, i.e, $G_o=D_4$ (see Appendix~\ref{appendix:hk_symmetries} for the specific details). If the pair-hopping is also absent, the combination of the continuous orbital $\text{U}(1)$ and the discrete orbital permutation symmetry $\mathcal{X}$ enlarges the symmetry to $G_o=\text{O}(2)$. Notably, there is two other points in the parameter space where, in the absence of $t_{\rm inter}$ and $\Delta_{\rm cf}$, the orbital symmetry group is continuous. The first has $G_o =O(2)$ and corresponds to $U^\prime=U-2J_H$ with $J_P=J_H$. In the second, the group corresponds to the fully rotationally invariant $\text{SU}(2)$ group at $J_P=0$ and $U^\prime=U-J_H$~\cite{Georges2013}.

These internal symmetries are also preserved by the non-interacting part of the Hamiltonian, which also allows for further discrete symmetries. First, the full Hamiltonian preserves time-reversal symmetry. The Kanamori interaction is inherently time-reversal invariant, and the kinetic terms are governed by purely real hopping amplitudes. Second, assuming strictly nearest-neighbor hopping on a bipartite lattice without crystal-field splitting, the system exhibits particle-hole symmetry at half-filling, fixing the chemical potential at $\mu^\ast=(U+2U^\prime-J_H)/2$.

\begin{figure*}[t]
    \centering
    \includegraphics[width=\linewidth]{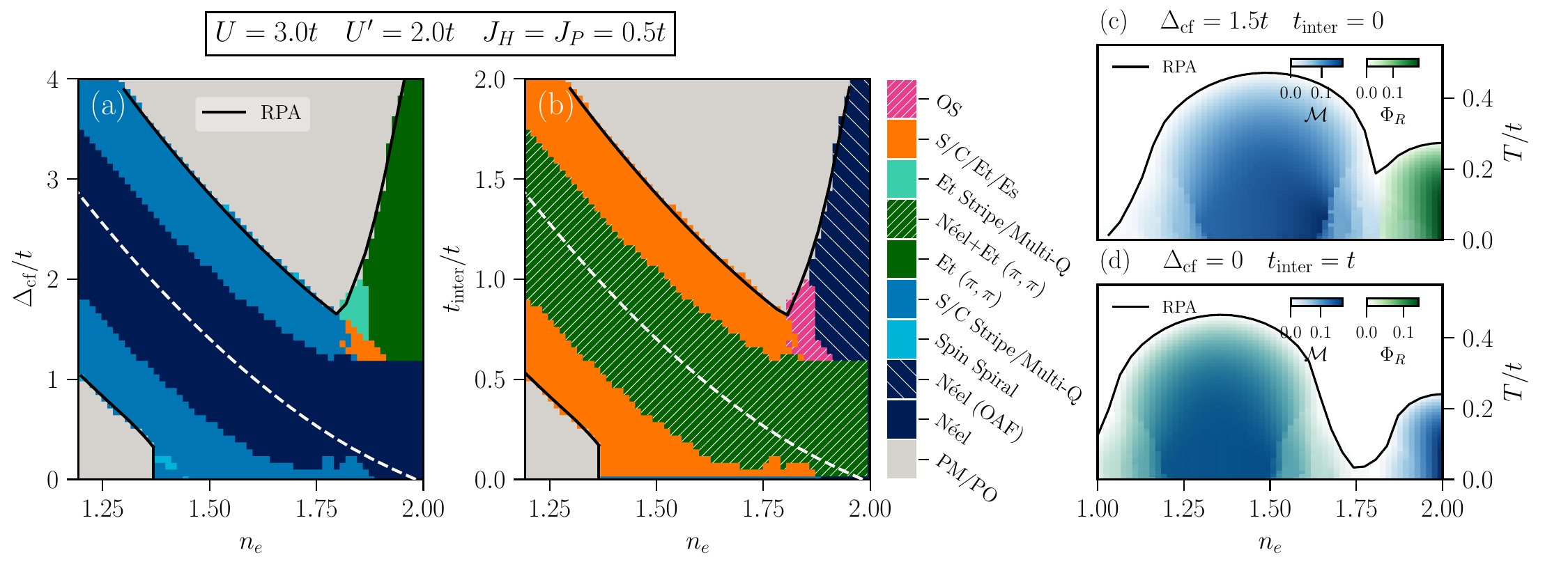}
    \caption{Effects of crystal-field splitting and inter-orbital hybridization. (a) Phase diagram as a function of crystal-field splitting $\Delta_{\rm cf}$ and electron density at a fixed temperature $T=0.1t$ for $t_{\rm inter}=0$. (b) Phase diagram as a function of inter-orbital hybridization $t_{\rm inter}$ and electron density at $T=0.1t$ for $\Delta_{\rm cf}=0$.
    The dashed white lines indicate the electron density of the non-interacting intra-band nesting condition as a function of $\Delta_{\rm cf}$ (a) and $t_{\rm inter}$ (b). Solid black lines are computed from the spin and charge RPA susceptibilities. 
    (c) Magnetic and real components of the $E_t$ order parameters as a function of temperature and electron density for $\Delta_{\rm cf}=1.5t$ and $t_{\rm inter}=0$. (d) Same as (c), but for a finite inter-orbital hybridization ($t_{\rm inter}=1.0t$) and $\Delta_{\rm cf}=0$.Other parameters: $L=20$, $U=3t$, $U^\prime=2t$, and $J_H=J_p=0.5t$.}
    \label{fig:delta_t_inter}
\end{figure*}

\section{Overview of the Multi-Orbital Orders}
\label{sec:orders_multi_orbital}

The extension from one to two orbitals allows for a much wider variety of ordered states due to the enriched symmetry group by the orbital degrees of freedom compared to those found in the single-orbital Hubbard model. In the repulsive single-orbital Hubbard model, which has been studied extensively within the Hartree-Fock approximation~\cite{Shraiman1989,chubukov1995,Igoshev2010,Igoshev2015}, ordering arises from an instability in the spin channel leading to the spontaneous symmetry breaking of the spin-SU(2) rotation symmetry, time-reversal symmetry and translation symmetry. However, it is important to note that for low-dimensional systems at finite temperatures, the spontaneous breaking of the continuous SU(2) symmetry is an artifact of the mean-field approximation, as it is strictly forbidden by the Mermin-Wagner theorem~\cite{PhysRevLett.17.1133}. As such, results of the Hartree-Fock approximation should only be regarded as guiding principles to possible electronic ordering tendencies. As a function of temperature and electronic density, the resulting phase diagram corresponds to different magnetic orders, which spontaneously break distinctively the translation symmetry group~\cite{PhysRevB.108.035139,PhysRevB.109.235149}. At half-filling, perfect Fermi-surface nesting~\cite{Fradkin_2013} drives the formation of local magnetic moments, which are ordered in a staggered pattern on the square lattice, the Néel antiferromagnetic state. Upon doping, the local magnetic order parameter acquires a different ordering vector that deviates from the $(\pi,\pi)$ point. In particular, spiral states emerge~\cite{PhysRevB.109.235149}, in which the local spin orientation rotates with a characteristic wave vector $\boldsymbol{Q}$, while the charge density remains uniform. At lower temperatures and higher doping levels, the system instead favors collinear spin- and charge-stripe order.

While on-site repulsion in the single-orbital Hubbard model drives purely magnetic instabilities, the Hubbard-Kanamori model incorporates Hund's coupling, inter-orbital density-density interactions, and pair hopping. This interplay tunes instabilities across both the spin and charge sectors, acting within either intra- or inter-orbital channels (\cref{fig:schematic}). When $\Delta_{\rm cf} = t_{\rm inter} = 0$, the Hamiltonian possesses an orbital $D_4$ symmetry that promotes four fundamentally distinct ordered phases: (a) conventional intra-orbital magnetic order, (b) Et order (i.e inter-orbital spin order), (c) intra-orbital charge order, and (d) Es order (i.e inter-orbital charge order). The temperature--density phase diagrams corresponding to representative regimes for each of these states are previewed in \cref{fig:T_vs_n_magnetic_dome_w_JP}(a)–(d).

For the Hamiltonian considered in this work, this orbital $D_4$ symmetry is lowered by a finite crystal-field splitting or an effective inter-orbital hybridization. In both cases, previously orthogonal order parameters are allowed to mix, reducing the number of distinct symmetry-broken states, as shown in the phase diagrams of \cref{fig:delta_t_inter}~(a) and (b). While the different states appearing across these phase diagrams are analyzed in detail in Secs.~\ref{sec:temp_doping_phase_diagrams} and \ref{sec:crystal_field_and_t_inter}, understanding their microscopic symmetry-breaking patterns and what distinguishes them requires first classifying the order parameters that define them.

Each ordered phase is identified by an observable transforming under a specific, non-trivial, irreducible representation (irrep) of the Hamiltonian symmetry group. Within the Hartree-Fock treatment, the effective mean-field Hamiltonian is quadratic, and all symmetry-breaking orders are encoded in the on-site one-body reduced density matrix, $\rho_j$~\footnote{Anomalous elements $\langle c_{j,\alpha,\sigma} c_{j,\alpha^\prime,\sigma^\prime}\rangle$ vanish as the mean-field Hamiltonian conserves total particle number.},
\begin{equation}
    \left(\rho_{j} \right)^{\alpha\alpha^\prime}_{\sigma \sigma^\prime} = \left\langle c^\dagger_{j,\alpha,\sigma} c^{\phantom{\dagger}}_{j,\alpha^\prime,\sigma^\prime} \right\rangle.
\end{equation}
To isolate the distinct symmetry channels, we expand $\rho_j$ in the complete Hermitian basis formed by the Kronecker product of orbital ($\tau$) and spin ($\sigma$) Pauli matrices:
\begin{equation}
    \rho_{j} = \sum_{a,b \in \{0,x,y,z\}} \mathcal{O}^{ab}_j \; \tau^a \otimes \sigma^b,
\end{equation}
where $\tau^0$ and $\sigma^0$ denote the $2 \times 2$ identity matrices. The expansion coefficients are given by
\begin{equation}
    \mathcal{O}^{ab}_j = \frac{1}{4} \sum_{\alpha,\gamma}\sum_{\mu,\nu} \tau^a_{\alpha\gamma} \sigma^b_{\mu\nu} \left\langle c^\dagger_{j,\alpha,\mu} c^{\phantom{\dagger}}_{j,\gamma,\nu} \right\rangle. 
    \label{eq:symmetry_breaking_decomposition}
\end{equation}

The representation spanned by the 16 observables $\mathcal{O}^{ab}_j$ can be decomposed into the irreps of the Hamiltonian's internal symmetry group, which constitute the respective order parameters, as summarized in \cref{tab:order_parameter_irreps}. In the following subsections, we systematically analyze these observables.

\begin{table}[t]
\caption{Classification of local bilinear observables by the irreducible representations of the symmetry group of the two-orbital Hubbard-Kanamori. The charge, spin, and time-reversal symmetries form the group $\mathcal{G}=\mathrm{U}_c(1) \times \mathrm{SU}(2) \rtimes \mathbb{Z}^T_2$, which acts independently of the orbital symmetries. All listed observables belong to the charge sector $C=0$ and are grouped by spin representation $S$. Table entries $(O, \mathcal{T})$ denote the orbital representation and time-reversal parity ($\mathcal{T}=\pm$). Under $D_4$, $O \in \{A_1, A_2, B_1, B_2\}$, whereas under $\mathbb{Z}^P_2$ (relative orbital parity) and $\mathbb{Z}^X_2$ (orbital exchange symmetry), $O = \pm$.\label{tab:order_parameter_irreps}} 
    \begin{ruledtabular}
        \begin{tabular}{l c c c}
            & \multicolumn{3}{c}{Symmetry Group} \\
            \cline{2-4} 
            & $\Delta_{\rm cf}=t_{\rm inter} = 0$ & $t_{\rm inter} = 0$ & $\Delta_{\rm cf} = 0$ \\
            \multicolumn{1}{c}{Observables}
            & $\mathcal{G}\times D_4$ 
            & $\mathcal{G}\times \mathbb{Z}^P_{2}$ 
            & $\mathcal{G}\times \mathbb{Z}^X_{2}$ \\
            \hline
            $S=0 \quad T^z$                        & $(B_2,+)$ & $(+,+)$\footnote{\label{footnote1} It is not an order parameter.} & $(-,+)$ \\
            $\phantom{S=0 \quad} \Psi_R$              & $(B_1,+)$ & $(-,+)$ & $(+,+)^{\ref{footnote1}}$ \\
            $\phantom{S=0 \quad} \Psi_I$              & $(A_2,-)$ & $(-,-)$ & $(-,-)$ \\
            \hline
            $S=1 \quad \boldsymbol{S}_{\rm sym}$              & $(A_1,-)$ & $(+,-)$ & $(+,-)$ \\
            $\phantom{S=1 \quad} \boldsymbol{S}_{\rm asym}$    & $(B_2,-)$ & $(+,-)$ & $(-,-)$ \\
            $\phantom{S=1 \quad} \boldsymbol{\Phi}_R$ & $(B_1,-)$ & $(-,-)$ & $(+,-)$ \\
            $\phantom{S=1 \quad} \boldsymbol{\Phi}_I$ & $(A_2,+)$ & $(-,+)$ & $(-,+)$ \\
        \end{tabular}
    \end{ruledtabular}
\end{table}

\subsection{Spin-Scalar Observables}
First, we examine the observables that transform as scalars under spin $\mathrm{SU}(2)$ rotations (the $S=0$ representation). The totally symmetric term ($a=0, b=0$) corresponds to the on-site particle number,
\begin{equation}
    \mathcal{O}^{00}_j = \frac{1}{4} \langle N_j \rangle = \frac{1}{4} \left( \langle n_{j,1} \rangle + \langle n_{j,2} \rangle \right).
\end{equation}
Regardless of the parameter regime, this term maps to the trivial irrep, remaining invariant under all symmetry operations. Next, combining the diagonal orbital matrix ($a=z$) with the spin identity ($b=0$) isolates the orbital polarization,
\begin{equation}
    \mathcal{O}^{z0}_j = \frac{1}{4} \left( \langle n_{j,2} \rangle - \langle n_{j,1} \rangle \right) = \frac{1}{2} \langle T^z_j \rangle.
\end{equation}
This operator flips sign under orbital permutation $\mathcal{X}T^z\mathcal{X}=-T^z$. As a result, it transforms as the odd representation of $\mathbb{Z}^X_2$ and the $B_2$ irrep of $D_4$. However, introducing a finite crystal field explicitly breaks this permutation symmetry, causing $T^z$ to revert to the trivial representation. While in \cref{fig:T_vs_n_magnetic_dome_w_JP}, a finite $\langle T^z \rangle$ points to a symmetry broken state in \cref{fig:delta_t_inter}~(a) $\langle T^z \rangle$ is trivially finite across the phase-diagram.

The remaining scalar observables arise from the off-diagonal orbital operators ($a \in \{x,y\}$). These describe singlet excitons. Combining them yields a complex local order parameter,
\begin{equation}
    \Psi_j = \mathcal{O}^{x0}_j + i\mathcal{O}^{y0}_j = \frac{1}{2} \sum_{\sigma} \langle c^\dagger_{j,1,\sigma} c^{\phantom{\dagger}}_{j,2,\sigma} \rangle.
\end{equation}
Its real ($\Psi_{R,j} = \mathcal{O}^{x0}_j$) and imaginary ($\Psi_{I,j} = \mathcal{O}^{y0}_j$) parts transform distinctly. While $\Psi_R$ is invariant under time-reversal and orbital exchange, $\Psi_I$ breaks both. Nevertheless, both components flip sign under relative orbital parity. Thus, within the $D_4$ group, $\Psi_R$ and $\Psi_I$ transform under the $B_1$ and $A_2$ irreps, respectively. If relative parity is explicitly broken, $\Psi_R$ transforms under the fully symmetric irrep. In this case, it ceases to be a true symmetry-breaking order parameter, as it happens in the phase diagram of \cref{fig:delta_t_inter}~(b), whereas $\Psi_I$ retains its non-trivial transformation.

Collectively, the orbital polarization with the excitonic terms form the local iso-orbital vector,
\begin{equation}
    \boldsymbol{T}_j = \frac{1}{2} \sum_{\gamma,\delta,\sigma} c^\dagger_{j,\gamma,\sigma} \boldsymbol{\tau}_{\gamma\delta} c^{\phantom{\dagger}}_{j,\delta,\sigma},
\end{equation}
which acts as the generators for the continuous orbital symmetries, at parameter points specified in the previous section.

\subsection{Spin-Vector Observables}

We now classify the observables transforming as vectors under spin-$\mathrm{SU}(2)$ rotations (the $S=1$ representation).

Coupling the diagonal orbital matrices ($a \in \{0,z\}$) with the spin vectors ($b \in \{x,y,z\}$) isolates the local intra-orbital spin densities. These form symmetric and antisymmetric combinations of the magnetic moments,
\begin{equation}
\begin{aligned}
\mathcal{O}^{0b}_j &= \langle \boldsymbol{S}_{j,{\rm sym}}^{b} \rangle = \frac{1}{2}\left( \langle \boldsymbol{S}_{j,2}^{b} \rangle + \langle \boldsymbol{S}_{j,1}^{b} \rangle \right)\\
\mathcal{O}^{zb}_j &=\langle \boldsymbol{S}_{j,{\rm asym}}^{b} \rangle= \frac{1}{2}\left( \langle \boldsymbol{S}_{j,2}^{b} \rangle - \langle \boldsymbol{S}_{j,1}^{b} \rangle \right).
\end{aligned}
\end{equation}
A non-zero expectation value here signals magnetic ordering, which fundamentally requires the spontaneous breaking of time-reversal symmetry. The symmetric component ($\boldsymbol{S}_{\rm sym}$) transforms trivially under the orbital group, indicating ferromagnetic alignment between the orbitals. In contrast, the antisymmetric component ($\boldsymbol{S}_{\rm asym}$) flips sign under orbital permutation, corresponding to an antiferromagnetic inter-orbital alignment (the $B_2$ irrep of $D_4$). Notably, both observables transform under the same irrep in the presence of a finite crystal field.

Finally, pairing the off-diagonal orbital operators ($a \in \{x,y\}$) with the spin matrices generates excitonic operators with a net spin polarization. These form the complex Et vector, accounting for the final six degrees of freedom and completing the basis of all $16$ local observables,
\begin{equation}
    \boldsymbol{\Phi}^{b}_j = \mathcal{O}^{xb}_j + i\mathcal{O}^{yb}_j = \frac{1}{2} \sum_{\mu,\nu} \langle c^\dagger_{j,1,\mu} \sigma^b_{\mu\nu} c^{\phantom{\dagger}}_{j,2,\nu} \rangle.
\end{equation}
Much like their singlet counterparts, the real ($\boldsymbol{\Phi}^b_{R,j} = \mathcal{O}^{xb}_j$) and imaginary ($\boldsymbol{\Phi}^b_{I,j} = \mathcal{O}^{yb}_j$) parts of this vector transform under the same irreps of the orbital symmetry group of $\Psi_R$ and $\Psi_I$. However, as they carry spin, their time-reversal parities are inverted, $\boldsymbol{\Phi}_{R}$ is odd, whereas $\boldsymbol{\Phi}_{I}$ is even under time-reversal.

\subsection{Space-Group Symmetries and Spatial Textures}

Besides inspecting the expectation values of local order parameters associated with internal symmetries, it is equally important to determine how these observables transform under the lattice space group. This leads to the different spatial textures, such as staggered, stripe~\cite{RevModPhys.75.1201,Xu2022,PhysRevB.109.235149}, and spiral phases~\cite{Shraiman1989,Igoshev2015,Bonetti2022} observed across the phase-diagrams of \cref{fig:T_vs_n_magnetic_dome_w_JP} and \cref{fig:delta_t_inter}. To achieve this, we analyze the Fourier transforms of the relevant local order parameters to extract their dominant ordering wave vectors, $\boldsymbol{Q}$. 

For scalar order parameters (denoted here generically as $\mathcal{O}_j$), the spatial modulation is captured by the Fourier transform,
\begin{equation}
    \mathcal{O}_{\boldsymbol{Q}} = \frac{1}{L}\sum_{j} \mathcal{O}_{j} e^{-i\boldsymbol{Q}\cdot \boldsymbol{r}_j}.
\end{equation}
By tracking the momentum $\boldsymbol{Q}$ that maximizes $\mathcal{O}_{\boldsymbol{Q}}$, we retrieve the translational symmetry-breaking pattern of the state.

For vector order parameters ($\boldsymbol{\mathcal{O}}_j$), we must additionally determine the internal directionality, namely, whether the vector field is collinear, coplanar, or non-coplanar. A collinear state is characterized by a vector field that fluctuates along a single global quantization axis, $\hat{\boldsymbol{e}}_1$. When a single unidirectional wave vector (such as $\boldsymbol{Q} = (Q, 0)$) governs this spatial modulation, it spontaneously breaks the lattice's $C_4$ rotational symmetry, resulting in a state known as a stripe.

In contrast, a coplanar state involves two orthogonal ordering directions, $\hat{\boldsymbol{e}}_1$ and $\hat{\boldsymbol{e}}_2$. A prominent example is the spiral state~\cite{Shraiman1989,Igoshev2015,PhysRevB.47.7910}, in which the magnitude of the vector field remains strictly uniform across the lattice, i.e, $|\boldsymbol{\mathcal{O}}_j| =\mathcal{O}_0$. This corresponds to a vector field that continuously rotates from site to site with a single pitch vector $\boldsymbol{Q}$,
\begin{equation}
    \boldsymbol{\mathcal{O}}_j = \mathcal{O}_0 \big[ \cos(\boldsymbol{Q}\cdot \boldsymbol{r}_j) \hat{\boldsymbol{e}}_1 + \sin(\boldsymbol{Q}\cdot \boldsymbol{r}_j) \hat{\boldsymbol{e}}_2 \big].
\end{equation}
Note that the standard staggered ordering emerges as a degenerate limit of this coplanar form, as when evaluated at $\boldsymbol{Q}=(\pi,\pi)$, the spiral state collapses into a collinear, alternating configuration.

Beyond single-$\boldsymbol{Q}$ modulations, the system can condense into multi-$\boldsymbol{Q}$ states formed by coherent superpositions of symmetry-related wave vectors. While single-$\boldsymbol{Q}$ stripes break rotational symmetry, symmetric multi-$\boldsymbol{Q}$ phases, such as a $2\boldsymbol{Q}$ checkerboard combining $(Q,0)$ and $(0,Q)$, preserve the underlying point-group while breaking the discrete translation symmetry. Notably, a finite momentum-grid resolution can also induce spurious multi-$\boldsymbol{Q}$ signatures, hindering the identification of the true ordering vector, even if this competing textures are thermodynamically disfavored~\cite{PhysRevB.108.035139}. In our phase diagrams, we therefore refrain from resolving the phase boundaries between these closely competing configurations, grouping them into broad phases. 

Naturally, such multi-$\boldsymbol{Q}$ or stripe textures can involve multiple order channels simultaneously, as exemplified by the coupled spin-charge stripes in the single-orbital Hubbard model. In the present case, the multi-orbital nature allows for the coexistence of even more exotic stripes that additionally incorporate excitonic degrees of freedom. A complete classification and labeling scheme for all these phases is detailed in Appendix~\ref{app:cat_orbital_components}.

\section{Temperature--Doping Phase Diagrams: $\Delta_{\rm cf}=t_{\rm inter}=0$}
\label{sec:temp_doping_phase_diagrams}

\begin{figure*}[t]
    \centering
    \includegraphics[width=\textwidth]{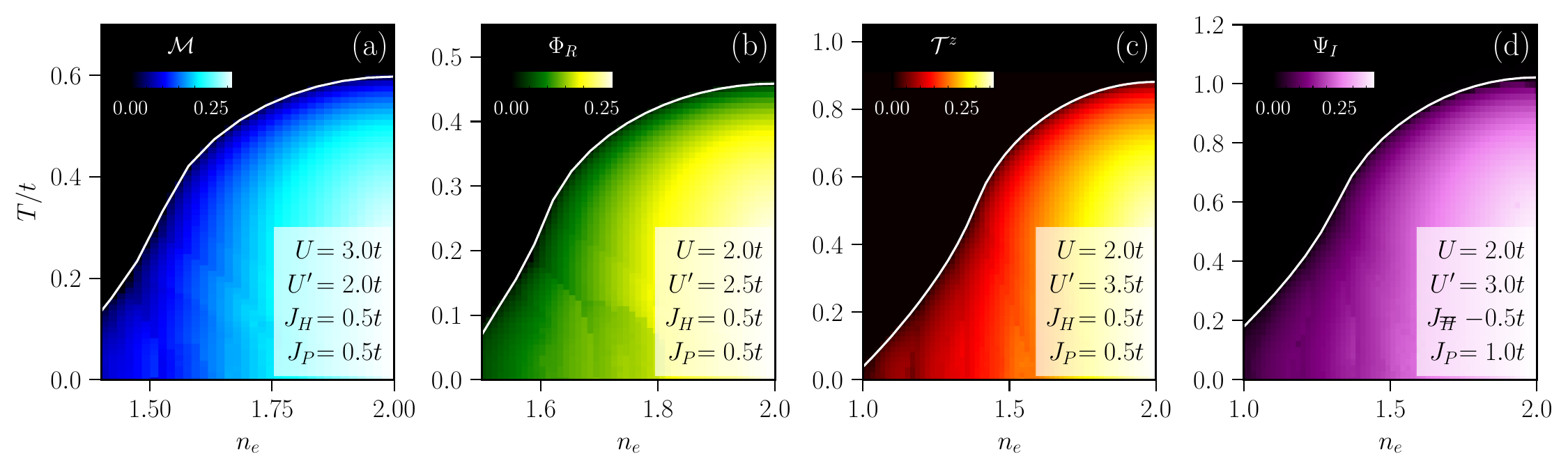}
    \caption{Average value of dominant order parameter as a function of the temperature and electronic density across the four main regimes. Panel (a) corresponds to the mainly magnetically ordered regime, the calculations are done with $U=3.0t$, $U^\prime=U-2J_H$. Panel (b) is representative of the excitonic ordered phase, the calculations used $U=2.0t$ and $U^\prime=2.5t$. Panel (c) shows the orbitally ordered region, where we used $U=2.0t$ and $U^\prime=3.5t$. Panel (d) corresponds to Es ordered regime, $U=2.0t$ and $U^\prime=3.0t$. The white lines correspond to the phase boundaries derived from the instabilities of the spin and charge susceptibilities in the paramagnetic and para-orbital state within the RPA. Other parameters: $L=20$, in panel (a), (b) and (c) we used $J_P=J_H=0.5t$ while in (d) $J_H=0.5t$ and $J_P=t$, and $\Delta_{\rm cf}=t_{\rm inter}=0$.}
    \label{fig:order_strenght}
\end{figure*}

We treat the interacting Hamiltonian within the Hartree-Fock approximation. In the real-space basis, this replaces the local four-body interaction term by a local, site-dependent one-body field determined by the matrix elements of the one-body reduced density matrix. These mean-field parameters are solved self-consistently without imposing a spatial ordering vector \textit{a priori}, allowing symmetry-broken phases, with any ordering vector commensurate with lattice size, and their boundaries to emerge naturally from the calculation (see Appendix~\ref{sec:HF_calculations} for details). As a check of the implementation, we explicitly derive the corresponding mean-field Hamiltonians at half-filling for the symmetry-broken states identified from the symmetry analysis. To characterize the onset of ordering from the high-temperature state, we further analyze its stability against fluctuations within the RPA~\cite{PhysRevB.75.224509,Graser_2009}. This analysis identifies the leading instability of the high-temperature unordered phase as a function of the model parameters and determines the corresponding critical temperature in the different regions of parameter space, complementing the Hartree-Fock calculation. All the unrestricted real-space Hartree-Fock calculations were performed on a $20\times 20$ lattice with periodic boundary conditions. Typically, we use at least ten different random initial configurations and at least $6000$ mean-field iterations to minimize the variational free energy. The mean-field parameters are typically converged to an absolute difference of $10^{-6}$; regions of poor convergence are marked as an unknown phase as detailed in Appendix~\ref{sec:HF_calculations}.

By starting from the high-temperature unordered phase, we determine the dominant symmetry-breaking channels by evaluating the spin ($\boldsymbol{\chi}^{zz}$) and charge ($\boldsymbol{\chi}^{\rho\rho}$) susceptibility tensors within the RPA~\cite{PhysRevB.75.224509,Graser_2009,PhysRevB.94.214515},
\begin{equation}
\begin{aligned}
\boldsymbol{\chi}^{zz}_{\mathrm{RPA}} (T,\boldsymbol{q},\omega) &= \left[\mathbb{I} - \hat{U}_S \boldsymbol{\chi}_0(T,\boldsymbol{q},\omega) \right]^{-1} \boldsymbol{\chi}_0(T,\boldsymbol{q},\omega), \\
\boldsymbol{\chi}^{\rho\rho}_{\mathrm{RPA}} (T,\boldsymbol{q},\omega) &= \left[\mathbb{I} + \hat{U}_C \boldsymbol{\chi}_0(T,\boldsymbol{q},\omega) \right]^{-1} \boldsymbol{\chi}_0(T,\boldsymbol{q},\omega),
\label{eq:rpa_equations}
\end{aligned}
\end{equation}
where $\boldsymbol{\chi}_0$ is the bare susceptibility tensor in the symmetric state satisfying $\boldsymbol{\chi}^{zz}_0 = \boldsymbol{\chi}^{\rho\rho}_0 = \boldsymbol{\chi}_0$, and $\hat{U}_S$ ($\hat{U}_C$) denotes the interaction vertex matrix in the spin (charge) channel associated with the Kanamori Hamiltonian (see Appendix~\ref{subsec:mean_field_paramagnetic}). A divergence in these susceptibilities signals an instability of the symmetric state, indicating the critical temperature $T^\star$, the instability wavevector $\boldsymbol{q}^\ast$, and the symmetry of the emergent order via the generalized Stoner condition:
\begin{equation}
    \det \left[ \mathbb{I} \mp \hat{U}_{S/C} \boldsymbol{\chi}_0(T^\star,\boldsymbol{q},0) \right] = 0.
    \label{eq:stonner_criterium_tc}
\end{equation}

In the limit $\Delta_{\mathrm{cf}} = t_{\mathrm{inter}} = 0$, the bare susceptibility reduces to that of decoupled, identical orbitals, rendering the susceptibility tensor diagonal and uniform in orbital space, $\boldsymbol{\chi}_0 = \chi_0 \mathbb{I}$. Here, the physical order is directly encoded in the orbital character of the instability: intra-orbital divergences in the spin (charge) sector indicate the onset of magnetic (charge-density-wave) order, whereas inter-orbital divergences signal triplet (singlet) excitonic condensation.

Under this condition, the matrix condition in Eq.~\eqref{eq:stonner_criterium_tc} factorizes into decoupled scalar equations of the form
\begin{equation}
    1 \mp \lambda \chi_0(T^\star, \boldsymbol{q}, 0) = 0,
    \label{eq:scalar_stoner}
\end{equation}
where $\lambda$ denotes the eigenvalues of $\hat{U}_{S}$ ($-$) or $\hat{U}_{C}$ ($+$). As the static susceptibility $\chi_0(T)$ grows monotonically upon cooling, the primary instability upon lowering temperature is governed exclusively by the largest positive eigenvalue across both sectors. Therefore, the phase boundaries separating competing symmetry-broken states are determined by the crossings of these leading eigenvalues, partitioning the interaction parameter space into four distinct regimes:
\begin{enumerate}
    \item Intra-orbital magnetic order, $U^\prime < U + J_H - |J_P|$;
     \item Orbital density waves, $U^\prime > U + |J_H| + |J_P|$;
    \item Triplet Excitonic order, $U + |J_H| - |J_P| < U^\prime < U + |J_H| + |J_P|$ and $J_H>0$;
    \item Singlet Excitonic order, $U - |J_H| - |J_P| < U^\prime < U + |J_P| + |J_H|$ and $J_H<0$;
\end{enumerate}
Notably, condensation near the critical temperature to a state with finite Et order requires a ferromagnetic Hund's coupling, while for the Es order, an antiferromagnetic Hund's coupling is needed instead.

Although our RPA calculation successfully identifies the dominant ordering channels near the critical temperature, it cannot capture transitions between ordered states or the filling-dependent behavior deep within each phase. Therefore, the following subsections map out the detailed temperature-doping phase diagrams for these four regimes.

\subsection{Magnetic Orders}

\begin{figure*}[t]
    \centering
    \includegraphics[width=\linewidth]{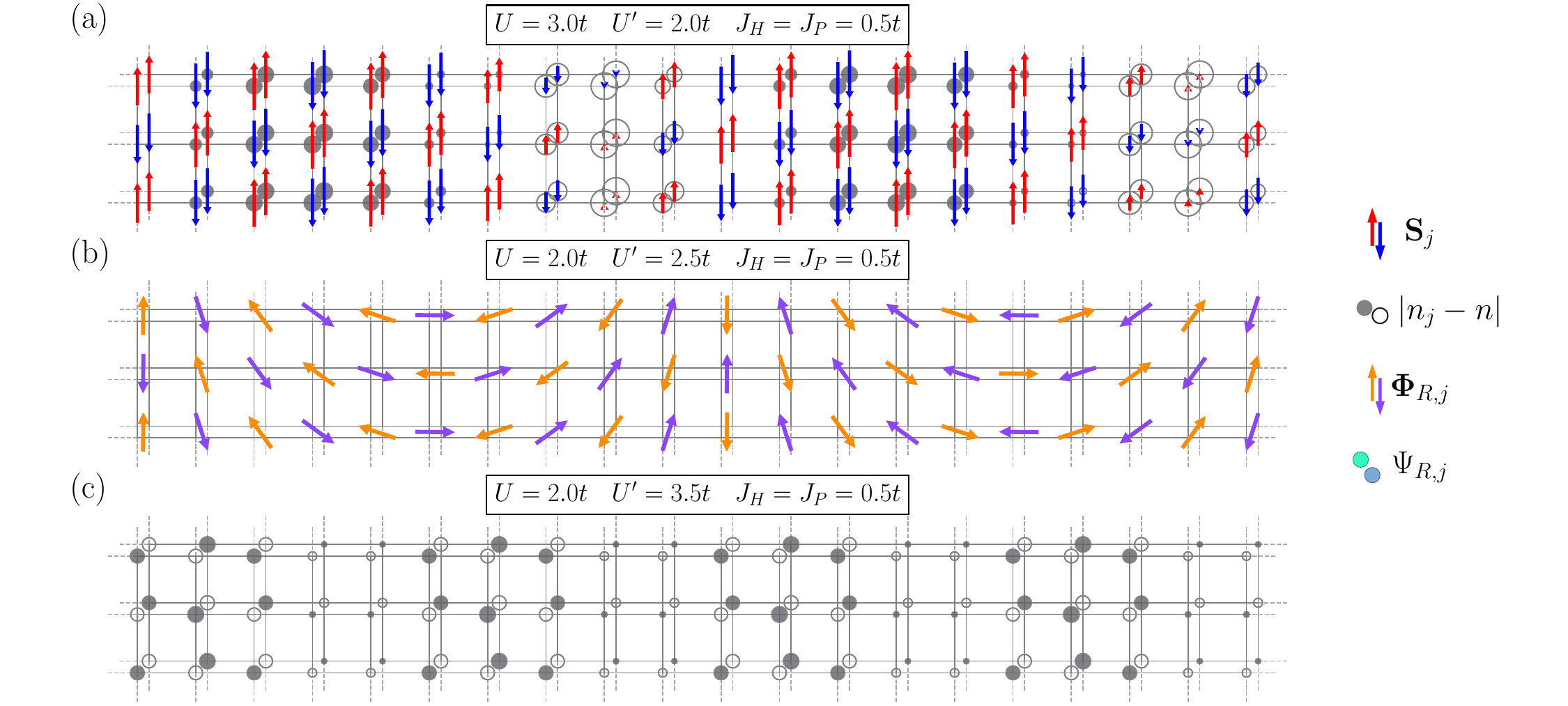}
    \caption{Representative orders of the doped Hubbard-Kanamori at finite temperature. (a) Spin-charge stripes. Charge and spin density waves, with orbitally aligned magnetic moments. ($U=3, U'=2,J_H=J_P=0.5$ with $n=1.81$, $T/t=0.055$). (b) Excitonic spiral. Orbital coherence leads to density waves in the spin singlet and triplet channels. These can coexist with the usual magnetic ordering, not plotted here ($U=2, U'=2.5,J_H=J_P=0.5$ with $n=1.81$, $T/t=0.085$).
    (c) Orbital density waves. A charge density wave forms in each orbital with a spatially modulated orbital imbalance ($U=2, U'=3.5,J_H=J_P=0.5$ with $n=1.51$, $T/t=0.31$). In (a) we plot the intra-orbital spin, (arrow color highlights the staggered magnetization) and charge density in each orbital, where an excess (deficit) of charge above (below) the average density is given by filled (empty) circles. In (c) there is no magnetic ordering only the charge density is shown. For (b), we only show the excitonic order parameters, but these can also coexist with magnetic ordering. The Es order is absence in the three plots. Other parameters: $L=20$ and $\Delta_{\rm cf}=t_{\rm inter}=0$.}
    \label{fig:real_space_order}
\end{figure*}

We consider $U^\prime < U + J_H - |J_P|$ with $J_H>0$, where the paramagnetic state has an instability in the intra-orbital spin channel. In particular, we set $U^\prime = U - 2J_H$, a parameter choice relevant to several classes of correlated materials~\cite{Kanamori1963,Georges2013}. In these systems, the low-energy physics is often described by an effective two-orbital Hubbard-Kanamori model, either capturing an active $e_g$ doublet or resulting from the active two-bands of the $t_{2g}$ manifold~\cite{Raghu2008,Luo2023,Igoshev2023,marino2026variationalmontecarlostudy}. While the relation $U^\prime = U - 2J_H$ reflects continuous rotational invariance in orbital space (which holds exactly for the $t_{2g}$ triplet), it is routinely retained in numerical studies of $e_g$ systems as a robust approximation.

In our numerical calculations, we used $U=3.0t$, $U'=2.0t$, and $J_H=J_P=0.5t$. This choice respects the expected hierarchy of energy scales predicted by \textit{ab initio} calculations for nickelate materials~\cite{Sakakibara2020,Kang2023} and keep the maximum interaction strength at a moderate value, since the static Hartree-Fock approximation tends to overestimate symmetry-breaking instabilities due to its omission of dynamical screening and quantum fluctuations. The relevant temperature--electron density phase diagram is shown in \cref{fig:T_vs_n_magnetic_dome_w_JP}~(a). Overall, this mean-field phase diagram closely resembles that of the single-orbital Hubbard model~\cite{PhysRevB.108.035139}. As expected from standard strong-coupling arguments, a robust Néel antiferromagnetic dome appears at half-filling and extends toward lower electron densities as the temperature increases. This stability is primarily driven by the intra-orbital Hubbard interaction and is further enhanced by Hund's coupling. These two mechanisms cooperate within the Hartree-Fock Hamiltonian, increasing the effective coupling to the Néel order parameter from $U$ to $U+J_H$, as we show in the Appendix~\ref{subsec:antiferromagnet_order} by doing the mean-field decoupling in this specific channel. No orbital-selective behavior is observed in this phase. Instead, each orbital independently develops a Néel antiferromagnetic order, while the local magnetic moments associated with different orbitals align ferromagnetically at each site due to Hund’s coupling, which favors parallel spin alignment between orbitals.

Upon hole doping (or, equivalently, electron doping because of the particle--hole symmetry), additional magnetic phases emerge with ordering vectors ($\boldsymbol{Q}$) that progressively deviate from the $(\pi,\pi)$ point. However, owing to the finite $20\times 20$ simulation cluster,only ordering vectors commensurate with the lattice can be resolved. As a result, some of the observed phases correspond to commensurate approximations of the true ordering vector and are stabilized by the finite momentum resolution, giving rise to slightly artificial states~\cite{PhysRevB.108.035139}. As such, some of the regions in the phase diagram shift to different orders in the thermodynamic limit. Similar to the single-orbital Hubbard model, there is the formation of a spiral phase at finite temperature: the charge is uniformly distributed across each orbital, but the intra-orbital magnetic moments form a non-collinear spiral: the local moments maintain a constant magnitude while their orientation rotates with pitch vector $\boldsymbol{Q}$. The Hund's coupling makes it highly energetically favorable for the two layers to have the exact same pitch vector, as the spins' moments remain aligned. This orbital-locked spin-spiral region appears in an extended region neighboring the Néel antiferromagnet phase, as observed in panel (a). At lower doping regions, there is a variety of orders where there are multiple dominant ordering vectors; this mostly occurs due to finite-size effects, as the optimal ordering wave vector is incommensurate in this lattice, and so is only truly resolved in larger clusters. As we also observe modulation of the charge in each orbital, these phases correspond to charge-spin stripe states. Notably, the charge-spin stripe forms with the same profiles in both orbitals due to the Hund's coupling and the presence of the orbital permutation symmetry $\mathcal{X}$. In \cref{fig:real_space_order}(a) we plot the real-space representative of this order. 

The phases discussed thus far are all characteristic of the single-orbital Hubbard model. However, a distinct regime emerges near the zero-temperature regime. In this region, we find that even though the Hund's coupling aligns the magnetic moments of the two layers, the magnetic strength differs in each layer, so that $\langle \boldsymbol{S}_{\rm asym} \rangle \neq 0$ and there is an orbital imbalance $\langle n_{j,1}\rangle \neq \langle n_{j,2}\rangle$. This phase is represented by the OS label. We note that even though, we observe this phase, this could be an artifact from the combination of the Hartree-Fock convergence and the optimal ordering vectors for these doping, therefore disappearing as we increase the system size.

In \cref{fig:order_strenght}~(a), we plot the average intra-orbital magnetization, 
\begin{equation}
    \mathcal{M} = \dfrac{1}{2L^2}\sum_{j,\alpha} \sqrt{\langle S^x_{j,\alpha} \rangle^2+\langle S^y_{j,\alpha} \rangle^2+\langle S^z_{j,\alpha} \rangle^2}.
    \label{eq:av_mag}
\end{equation}
The magnetic order is strongest in the N\'eel antiferromagnet phase and is suppressed in regions where spiral or stripe orders emerge. The average magnetization evolves continuously from the paramagnetic to the ordered phase. Consistent with this continuous transition, our analytical derivation at half-filling (see Appendix~\ref{subsec:antiferromagnet_order}) demonstrates that the N\'eel order parameter exhibits the mean-field scaling $\propto |T-T_c|^{1/2}$ near the critical temperature $T_c$. The critical temperature depends on both the repulsive Hubbard interaction and Hund's coupling, and increases with the Hund's coupling across all values of doping. This is consistent with the formation of stronger local magnetic moments for increasing $J_H$.

\subsection{Triplet Excitonic Orders}
\begin{figure}[t]
    \centering
    \includegraphics[width=\linewidth]{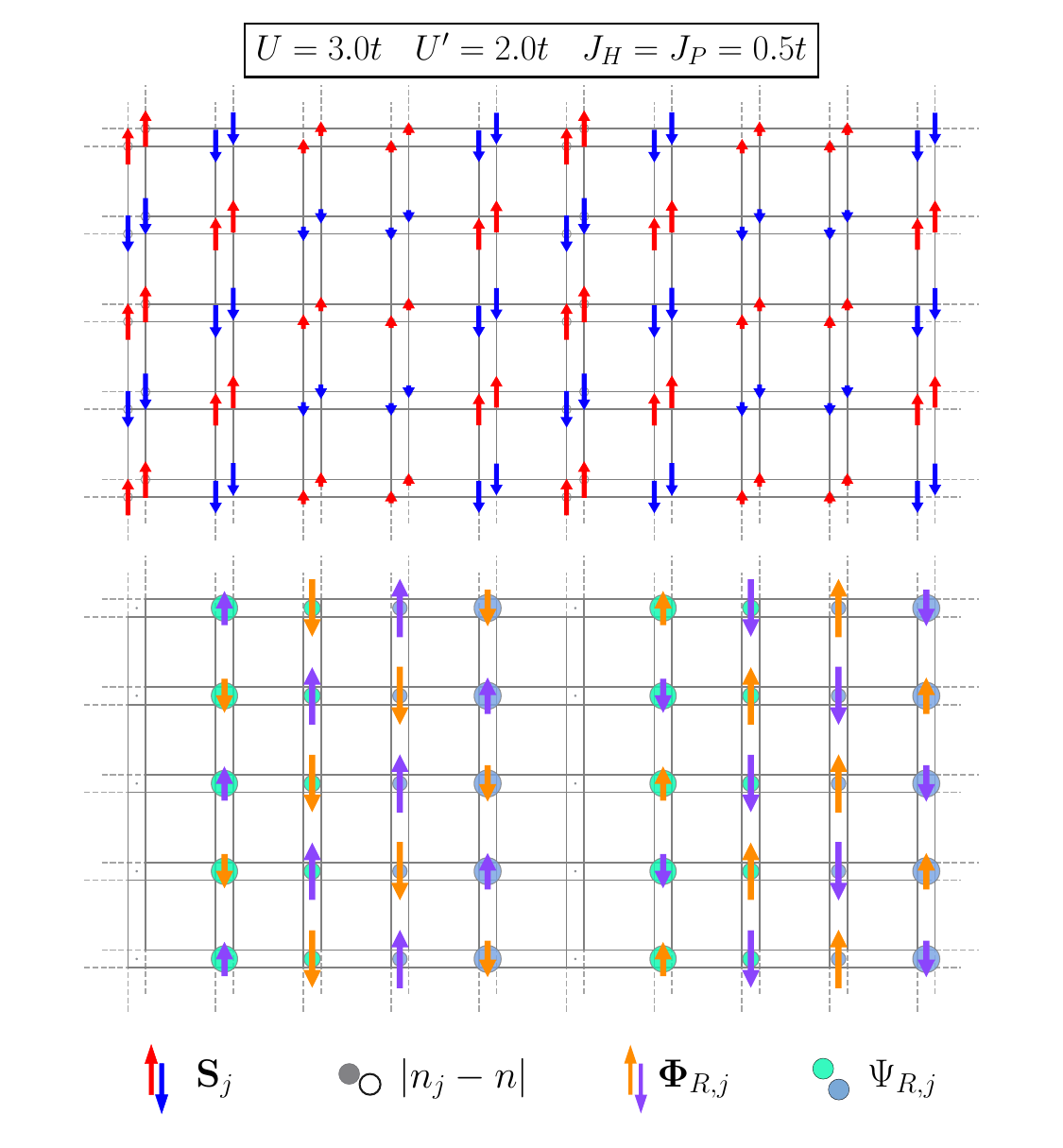}
    \caption{
    Example of coexisting magnetic and excitonic stripe order in the Hubbard-Kanamori model. Top shows the magnetic and charge ordering (in this case homogeneous charge density) and bottom shows the excitonic ordering. Triplet order is shown with arrows, whilst singlet order is denoted by circles - green (blue) corresponds to a positive (negative) value of Es. 
    Parameters: $U=2t, U'=2.5t, J_H=J_P=0.5t$ at a temperature of $T=0.042t$ and density $n=1.6$ with $L=20$ and $\Delta_{\rm cf}=t_{\rm inter}=0$.}
    \label{fig:real_space_stripe_excitonic}
\end{figure}

\begin{figure}
    \centering
\includegraphics{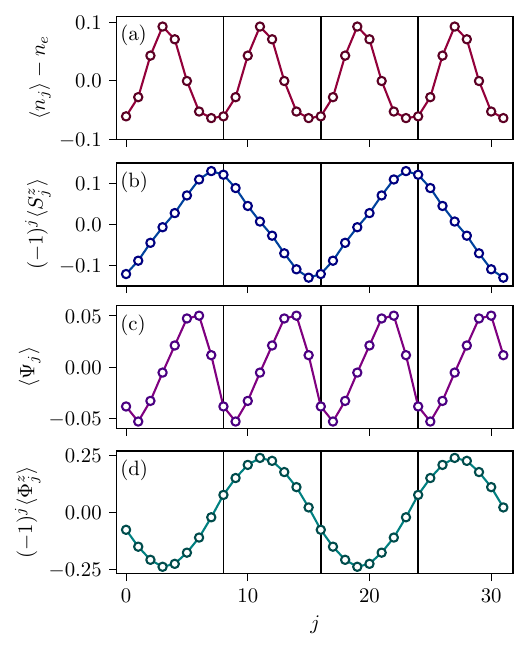}
    \caption{One-dimensional profiles of the (a) charge density, (b) intra-orbital magnetization, and excitonic (singlet (c) and triplet (d)) order along the longitudinal $x$ direction. Calculations are performed on a lattice with $L=32$ and width of $W=8$ sites at temperature $T=0.01t$, using interaction parameters $U=2t$, $U^\prime=2.5t$, and $J_H=J_P=0.5t$. The system is hole-doped at $x=1/8$ away from half-filling (average electron density $n_e=1.75$). At this doping level, the system stabilizes a stripe phase characterized by a charge and Es wavelength of $\lambda_c = 8$ lattice sites, accompanied by spin and Et vector fields that modulate with a wavelength of $\lambda_s = 16$ lattice sites.}
    \label{fig:stripe_profiles}
\end{figure}

Increasing $U^\prime/U$ reveals an additional instability driven by the Hubbard-Kanamori interactions, stabilizing an intermediate phase in the regime $U + |J_H| - |J_P| < U^\prime < U + |J_H| + |J_P|$, \cref{fig:T_vs_n_magnetic_dome_w_JP}(b). Here, the leading instability near the critical temperature remains magnetic, but shifts to the inter-orbital channel. This transition spontaneously breaks both the continuous $\mathrm{SU}(2)$ spin-rotation symmetry and the relative orbital parity symmetry defined in \cref{eq:relative_parity} (see \cref{tab:order_parameter_irreps}), which produces a finite Et vector order parameter, $\boldsymbol{\Phi}_j$.

The real and imaginary components of $\boldsymbol{\Phi}_j$ correspond to distinct physical phases, we determine the favored one near $T_c$ by projecting the interactions into the Et channels,
\begin{equation}
   H = - \frac{1}{2} \sum_{\boldsymbol{k}} \left[ (U^\prime + J_P) \mathbf{\Phi}_R^2(\boldsymbol{k}) + (U^\prime - J_P) \mathbf{\Phi}_I^2(\boldsymbol{k}) \right].
   \label{eq:Heff_diagonalized}
\end{equation}
The effective interaction vertices governing the Stoner divergence of the static spin susceptibility (\cref{eq:stonner_criterium_tc}) are $V_R = U^\prime + J_P$ and $V_I = U^\prime - J_P$. For $J_P > 0$, $V_R > V_I$, which marks the real channel as the dominant instability. The system therefore condenses into a purely real Et state near $T_c$, consistent with our mean-field analysis at half-filling (Appendix~\ref{appendix:mean_field_anzats_rpa}).

While this instability analysis dictates the behavior near $T_c$, it does not preclude an imaginary component from developing deeper within the ordered phase. To test this, we evaluate the spatial order-parameter magnitudes,
\begin{equation}
   \Phi_{R/I} = \frac{1}{L^2} \sum_{j} |\boldsymbol{\Phi}_{R/I,j}|,
   \label{eq:triplet_order_param}
\end{equation}
across the phase diagram. We find that $\Phi_I = 0$ and $\Phi_R$ is finite throughout the region shown in \cref{fig:order_strenght}(b). Consequently, the state spontaneously breaks time-reversal symmetry while preserving the orbital-exchange symmetry.

The phase diagram in \cref{fig:T_vs_n_magnetic_dome_w_JP}(b) reveals several extended phases featuring pure or intertwined Et density waves. At half-filling, a commensurate Et density wave emerges with ordering vector $\boldsymbol{Q} = (\pi,\pi)$, governed by the peak in the non-interacting spin susceptibility. Although this state is characterized by a finite inter-orbital spin-triplet order parameter, each orbital remains locally paramagnetic $\langle S_{j,\alpha} \rangle = 0$ and equally occupied. Hole doping shifts the ordering to an incommensurate wavevector $\boldsymbol{Q}$, analogous to magnetic charge-stripe formation. In real space, this incommensurate state exhibits spatial amplitude modulations of the triplet vector coexisting with Es order, \cref{fig:real_space_stripe_excitonic}. In \cref{fig:stripe_profiles} the modulation of the charge, spin and excitonic orders along the $x$ direction is shown for a system hole doped at $1/8$ away from half-filling and $T=0.01t$. This points lies within the orange region in \cref{fig:T_vs_n_magnetic_dome_w_JP}(b). We observe that, much like the spin-charge stripes in the single-orbital Hubbard model, the charge profile exhibits a wavelength inversely proportional to the doping level (8 sites in the present case), with the intra-orbital magnetization profile having a wavelength twice as long. Notably, the Es and Et stripes adhere to this same scaling behavior, having wavelengths of 8 and 16 sites, respectively. At finite temperatures, the system realizes a spiral Et state, \cref{fig:real_space_order}(b), in which the magnitude of $\langle \boldsymbol{\Phi}_j \rangle$ is spatially uniform while its orientation rotates with a well-defined pitch vector $\boldsymbol{Q}$,
\begin{equation}
    \langle \boldsymbol{\Phi}_j \rangle = \langle \boldsymbol{\Phi}_0 \rangle \left[\cos(\boldsymbol{Q}\cdot \boldsymbol{r}_j)  \hat{e}_1 + \sin(\boldsymbol{Q}\cdot \boldsymbol{r}_j)  \hat{e}_2 \right],
\end{equation}
where $\langle \boldsymbol{\Phi}_0 \rangle$ is the amplitude of the excitonic order parameter, $\hat{e}_1$ and $\hat{e}_2$ are two orthonormal unit vectors. Upon cooling, this spiral state gives way to an extended region where the Et pattern intertwines with intra-orbital spin and charge stripes, \cref{fig:real_space_stripe_excitonic}.

Pair hopping plays an indispensable role in stabilizing the Et phase. In the absence of $J_P$, RPA calculations indicate that near the critical temperature, the leading instability shifts from the intra-orbital spin sector to the intra-orbital charge sector. In this limit at half-filling, our numerical results show a direct transition at $U^\prime = U + J_H$ from a N\'eel antiferromagnet to a staggered orbital density wave (ODW). A finite $J_P$ directly reinforces the mean-field excitonic decoupling channel, scaling the mean-field gap as $U^\prime + J_P$ (see Appendix~\ref{appendix:mean_field_anzats_rpa}). Furthermore, a non-zero $J_P$ reduces the orbital symmetry group, as the excitonic phase breaks only a discrete relative orbital parity symmetry rather than the continuous $\mathrm{U}(1)$ orbital symmetry associated with orbital occupation imbalance.

\subsection{Orbital Orders}

\begin{figure}[t]
\centering
\includegraphics{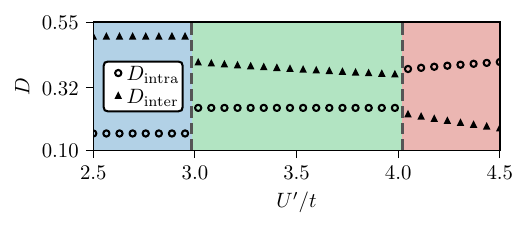}
    \caption{Mean inter- and intra-orbital double occupancy as a function of $U^\prime/t$ at half-filling in the ordered phases. The dashed lines show the critical point separating the magnetically (blue), Et (green) and orbitally (red) ordered phases. Other parameters: $L=20$, $U=3.0t$, $J_H=J_P=0.5t$ and $T=0.1t$.}
    \label{fig:double_occupancy}
\end{figure}

In the parameter regimes discussed thus far, orbital ordering is largely suppressed by the combined effects of the Hund's coupling $J_H$ and intra-orbital Hubbard repulsion $U$, which favor high-spin, magnetic states. This behavior changes drastically when the inter-orbital repulsion $U^\prime$ becomes the dominant energy scale ($U^\prime > U + J_H + J_P$). Here, the primary instability shifts to the charge sector: to minimize the severe inter-orbital energy penalty, the system spontaneously breaks the local $\mathbb{Z}_2$ orbital permutation symmetry by preferentially forming intra-orbital doublons. In a strong-coupling expansion, the associated energy scales as $E \propto -U^\prime\sum_j( T^z_j)^2$. This transition is evident in \cref{fig:double_occupancy}, where we compute the intra- and inter-orbital double occupancy, 
\begin{equation}
\begin{aligned}
    D_{\rm intra} = \dfrac{1}{2 L^2}\sum_{j,\alpha} \langle n_{j,\alpha,\uparrow} n_{j,\alpha,\downarrow} \rangle,\\
    D_{\rm inter} = \dfrac{1}{2 L^2}\sum_{j,\sigma} \langle n_{j,1,\sigma} n_{j,2,\sigma} \rangle.
\end{aligned}
\end{equation}
While intra-orbital doublons are suppressed in the magnetic and excitonic phases, this large $U^\prime$ regime is characterized by a stark increase in intra-orbital double occupancy and a corresponding suppression of inter-orbital doublons. Because Pauli exclusion restricts these intra-orbital pairs to a spin-singlet state, the local magnetic moment $\langle \boldsymbol{S}_{i\alpha} \rangle$ for site $i$ and orbital $\alpha$ is completely quenched at the mean-field level ($\langle \boldsymbol{S}_{i\alpha} \rangle = 0$)~\cite{Capone2026}.

The resulting phase diagram is shown in \cref{fig:T_vs_n_magnetic_dome_w_JP}(c), where three main regions emerge. Around half-filling and at higher temperatures, a spatially staggered ODW forms. In this phase, the electronic density localizes predominantly in one orbital at a given site, but alternates to the opposite orbital on neighboring sites to minimize the hopping energy. In the Appendix~\ref{appendx:mf_anzats_odw}, we derive the mean-field \textit{Ansatz} for this order, from where we find that the mean-field gap increases as a function of $2U^\prime - U - J_H$. In contrast to the magnetic regime, the Hund's coupling is detrimental to it.

Upon hole-doping, the ordering vector shifts away from $(\pi,\pi)$, creating a spatially modulated ODW, \cref{fig:real_space_order}(c), that continues to avoid inter-orbital double occupancy. Due to the robust formation of local spin-singlets, these ODW phases generally exhibit no signs of magnetic or Et order. Finally, as shown in \cref{fig:order_strenght}(c), the total orbital imbalance,
\begin{equation}
    \mathcal{T}^z = \dfrac{1}{2L^2}\sum_{j,\sigma} \left|\left\langle T^z_{j,\sigma} \right\rangle \right|,
\end{equation}
is maximized by the ODW ordered at $(\pi,\pi)$ and decreases at lower fillings.

\subsection{Singlet Excitonic Phases}
\begin{figure*}[t]
    \centering
    \includegraphics[width=\linewidth]{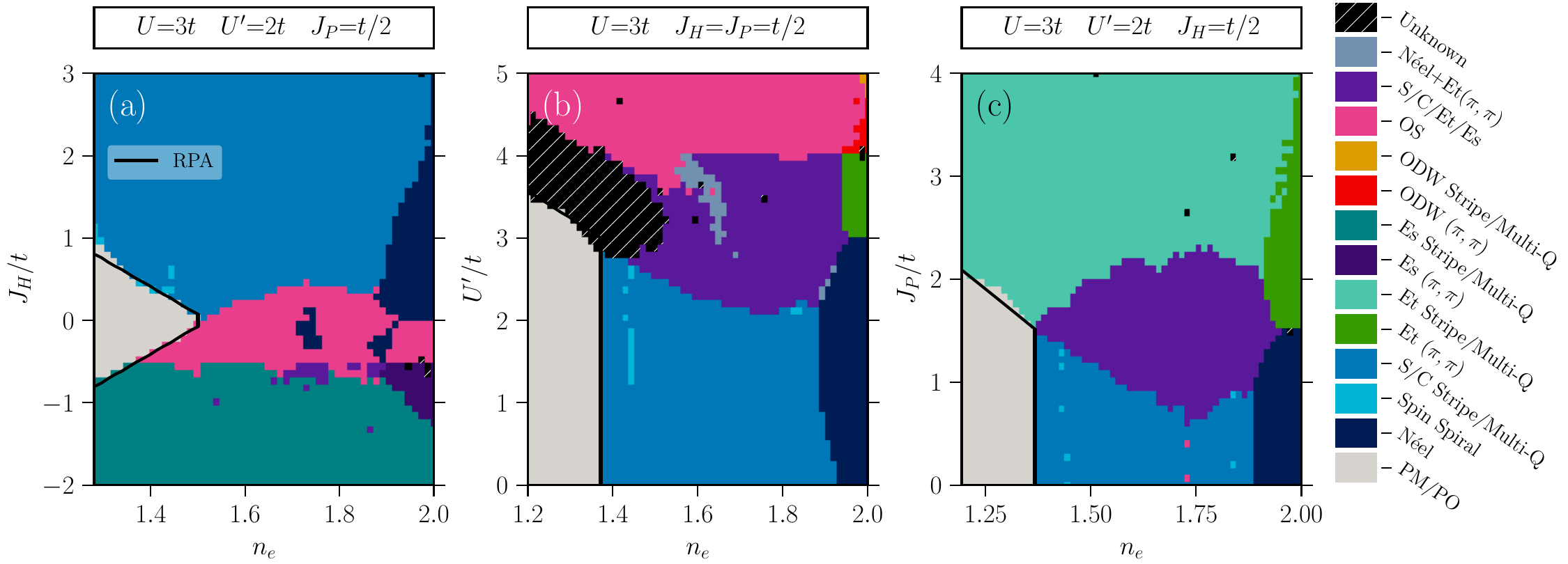}
    \caption{Low-temperature mean-field phase diagrams of the Hubbard-Kanamori model, varying each of the inter-orbital interaction terms in turn. Starting from the rotationally symmetric point ($U=3t$, $J_H=J_P=0.5t$, $U'=U-2J_H$), we vary (a) $J_H/t$, (b) $U'/t$, and (c) $J_P/t$ against electron density, while holding the other parameters constant. The black line corresponds to the phase boundaries derived from the RPA spin/charge susceptibilities. Other parameters: $L=20$, $T=0.1t$ and $\Delta_{\rm cf}=t_{\rm inter}=0$.}
    \label{fig:parameter_scans}
\end{figure*}
The final instability to be discussed leads to the formation of an Es dome. Similarly to the orbital density-wave phase, this instability occurs within the charge channel, but, in contrast, it emerges in the inter-orbital rather than the intra-orbital sector. As derived from the RPA condition and verified at half-filling in the Appendix~\ref{appendix:mean_field_anzats_rpa} using an explicitly constrained \emph{Ansatz}, the emergence of this phase requires a negative Hund's coupling and substantial inter-orbital repulsion. Importantly, this inter-orbital repulsion cannot be overwhelmingly large, as the orbital density-wave phase would otherwise become energetically more favorable. Physically, a negative Hund's coupling favors a local spin-singlet state over a high-spin configuration. In \cref{fig:T_vs_n_magnetic_dome_w_JP}(d), we present the temperature-doping phase diagram for the parameter set $U=2.0t$, $U^\prime=3.0t$, $J_H=-0.5t$, and $J_P=1.0t$ and in panel (d) of \cref{fig:order_strenght} the respective total strength of the imaginary singlet order parameter is plotted, 
\begin{equation}
    \Psi_{R/I} = \dfrac{1}{L^2}\sum_j \left|\Psi_{R/I,j}\right|.
\end{equation}
Similar to the Et phase, the Es order parameter can either condensed into a purely real or imaginary value. Near the critical temperature, the RPA calculation indicates that the imaginary component condenses with a higher critical temperature when $J_P>0$, while the real component is preferred for $J_P<0$. This is again derived by obtaining the interaction vertices by expressing the Kanamori interaction in terms of these channels,
\begin{equation}
    H = -\dfrac{1}{2} \sum_{\boldsymbol{k}} V_R \Psi^2_R(\boldsymbol{k}) + V_I\Psi^2_I(\boldsymbol{k}), 
\end{equation}
where $V_R=\left(U^\prime -2J_H - J_P \right)$ and $V_I=\left(U^\prime -2J_H + J_P \right)$. We obtain numerically $\Psi_R=0$ across the entire temperature-doping phase diagram, which indicates that this Es phase, besides breaking the relative orbital parity, also lacks the orbital permutation and time-reversal symmetry. Consequently, beyond mean-field theory, these states can potentially persist at finite temperatures in two dimensions without violating the Mermin-Wagner theorem~\cite{PhysRevLett.17.1133}.

Exactly at half-filling, we observe the formation of a staggered Es density wave. This occurs because charge susceptibility exhibits an instability at $\boldsymbol{Q}=(\pi,\pi)$ due to perfect Fermi-surface nesting. As we show analytically in the Appendix~\ref{appendix:mean_field_anzats_rpa}, the order parameter condenses and remains purely imaginary till $T=0$. The resulting Es order indicates that there is a spontaneous, staggered hybridization between the two orbitals. In contrast to the ODW phase, breaking the orbital permutation symmetry does not lead to a spatial imbalance of the charge within each orbital, this remains spatially uniform and equal for both orbitals $\langle n_1 \rangle = \langle n_2\rangle$. Upon hole doping, the magnitude of the Es order decreases and the favorable ordering vector moves away from the $(\pi,\pi)$ point in the first Brillouin zone. In this regime, the system forms Es stripes. Notably, there is a region in the phase diagram around $T\sim0.2t$ in the vicinity of the transition from a staggered singlet density wave order into a stripe one, where the system also forms orbital density-waves.

\subsection{Robustness of the Regimes: Varying $J_H$, $U'$, and $J_P$}
\label{sec:parameter_scans}

While the RPA analysis identifies the normal-state instabilities near the critical temperature, it leaves the nature of the ground state and secondary transitions within the ordered regime unresolved. To address this and explore the landscape of competing and intertwined orders, we examine the Hartree-Fock phase diagrams as a function of electronic density and the different interaction terms. Our baseline is the material-relevant reference point of \cref{fig:T_vs_n_magnetic_dome_w_JP}~(a), defined by $U=3.0t$, $U'=2.0t$, and $J_H=J_P=0.5t$.
To isolate the effect of individual interactions, we systematically vary the Hund's coupling $J_H$, the inter-orbital Coulomb repulsion $U'$, and the pair-hopping amplitude $J_P$ around this baseline while keeping the remaining parameters fixed. The resulting phase diagrams as a function of doping are presented in \cref{fig:parameter_scans}(a)--(c). All calculations are carried out at $T=0.1t$, which ensures reliable numerical convergence and faithfully approximates the low-temperature physics.

\Cref{fig:parameter_scans}~(a) illustrates the role of Hund's coupling. For ferromagnetic coupling ($J_H > 0$) with $U' < U$, the leading instability at half-filling occupies the intra-orbital spin channel, stabilizing a conventional N{\'e}el antiferromagnet. As $J_H$ decreases, the system transitions directly from the N{\'e}el antiferromagnet into an orbital-selective magnetic phase characterized by unequal local moments, $\langle \boldsymbol{S}_{j,1} \rangle \neq \langle \boldsymbol{S}_{j,2} \rangle$, before eventually forming a Es density wave with ordering vector $\mathbf{Q} = (\pi,\pi)$. Upon finite hole doping, the $J_H > 0$ regime is predominantly governed by coupled spin-charge stripe phases. At sufficiently strong ferromagnetic Hund's coupling, these stripes remain mutually aligned across both orbitals. Conversely, strong antiferromagnetic coupling ($J_H < 0$) destabilizes the stripe order in favor of a robust Es condensate.

At half-filling, increasing the inter-orbital interaction $U'$ drives successive direct transitions from the N{\'e}el antiferromagnet to a staggered Et density wave, and subsequently to a staggered ODW, with phase boundaries consistent with the RPA analysis of the previous section. Upon finite hole doping, in addition to the formation of conventional spin-charge stripes for $U' < U + J_H - J_P$, we observe an extended coexistence region where both the intra-orbital magnetic and Et order parameters develop finite expectation values.

Finally, increasing the pair-hopping interaction $J_P$ promotes the formation of Et order, as shown in \cref{fig:parameter_scans}(c). At half-filling, the transition from the N{\'e}el state to a staggered Et phase occurs once $J_P > U - U' + J_H$. With finite doping, the system generally develops incommensurate charge-modulated states: enhancing $J_P$ drives the system from spin-charge stripes to excitonic stripe order via an intermediate coexistence regime where both symmetries are simultaneously broken. From a symmetry perspective, the onset of Et order is closely analogous to magnetic ordering, as both spontaneously break spin $\mathrm{SU}(2)$ and time-reversal symmetry. However, stabilizing an Et condensate additionally requires breaking the relative orbital parity symmetry.

\section{The Role of Crystal Field Splitting and Inter-orbital Hybridization}
\label{sec:crystal_field_and_t_inter}

We now discuss in detail the phase diagram of \cref{fig:delta_t_inter}~(a). The phase diagrams are obtained for the regime $U^\prime = U - 2J_H$ with $J_P=J_H$. We consider the same parameters as before: $U=3.0t,U^\prime=2.0t,J_H=J_P=0.5t$, where the system develops intra-orbital magnetic order due to dominance of the intra-orbital Hubbard interaction. A finite crystal-field splitting directly competes with the intra-orbital Hubbard repulsion. While $\Delta_{\rm cf}$ favors double occupancy of the lower-energy orbital, thereby enhancing the local magnitude of $T^z$, the Hubbard repulsion penalizes such configurations and instead favors a more balanced orbital occupation with $\langle T_j^z\rangle=0$. At half-filling, as one increases $\Delta_{\rm cf}$, the system undergoes a transition from a Néel antiferromagnetic phase with vanishing orbital polarization to an Et phase with a purely real component. For large enough crystal-field splitting, the Et phase is eventually superseded by an orbitally polarized band insulator.

\begin{figure}[t]
    \centering
    \includegraphics{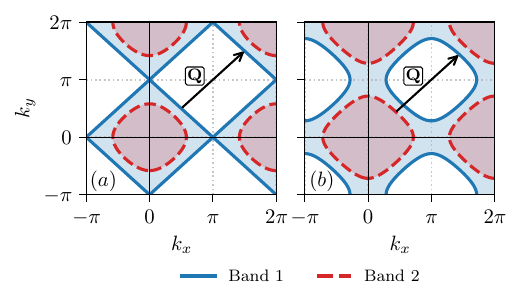}
    \caption{Fermi surface nesting conditions in the presence of crystal-field splitting. (a) Intra-orbital nesting condition corresponding to density $n^\star_{-}$, here $\mu_{\rm eff}=-\tilde{\Delta}_{\rm cf}/2$ or ($\mu_{\rm eff}=-\tilde{t}_{\rm inter}/2$). Perfect nesting within the lower-energy orbital (Band 1) drives the strongest divergence in the intra-orbital susceptibility $\left[\chi^{\rm intra}_0 (\boldsymbol{q},\omega)\right]^{1,1}_{1,1}$ and $\left[\chi^{\rm intra}_0 (\boldsymbol{q},\omega)\right]^{2,2}_{2,2}$, promoting the formation of the magnetic dome. (b) Inter-orbital nesting condition at half-filling ($\mu_{\rm eff} = 0$). The fundamental nesting vector $\mathbf{Q}=(\pi,\pi)$ perfectly connects the macroscopic occupied states ($E < E_F$, shaded regions) of the lower-energy orbital with the unoccupied states of the higher-energy orbital (Band 2, dashed contours).}
    \label{fig:fs_instabilities_crystal_field}
\end{figure}

As one dopes the system away from half-filling, the system transitions out of the staggered Et phase, and, for sufficient doping, enters a dome of Néel order. The Et and intra-orbital magnetic domes are seen in the temperature-doping plots in \cref{fig:delta_t_inter}~(c), where we plot the average magnetization (\cref{eq:av_mag}) and the real component of the average Et order (\cref{eq:triplet_order_param}). At the periphery of the magnetic dome, one finds the usual charge-spin stripes known to the single-orbital Hubbard model. Before discussing the evolution of the phase boundaries, we first identify the microscopic origin of the magnetic and excitonic domes. As evident in \cref{fig:delta_t_inter}~(a), the Et phase remains centered at half-filling, whereas the antiferromagnet dome shifts toward lower electron densities as $\Delta_{\rm cf}$ increases. This behavior can already be understood from the high-temperature paramagnetic phase (see Appendix~\ref{subsec:mean_field_paramagnetic}). A finite crystal-field splitting increases the orbital degeneracy, shifting the two non-interacting bands by $\pm\Delta_{\rm cf}/2$. Consequently, the chemical potential at which perfect Fermi-surface nesting occurs is displaced away from half-filling. The corresponding electron densities are entirely determined by the crystal-field splitting as these are given above $(+)$ and below $(-)$ half-filling as
\begin{equation}
    n^\star_{\pm} = \dfrac{2}{L^2}\sum_{\boldsymbol{k},\lambda = \pm}f_{\rm FD} \left(\varepsilon_{\boldsymbol{k}} +\frac{\tilde{\Delta}_{\rm cf}}{2} (\lambda\pm1)\right),
    \label{eq:dome_density_cf}
\end{equation}
with $f_{\rm FD}\left( \cdot \right)$ the Fermi-Dirac function and $\tilde{\Delta}_{\rm cf}$ the crystal field splitting with the Hartree correction (see Appendix~\ref{subsec:mean_field_paramagnetic}). In \cref{fig:delta_t_inter}~(a), the  dashed white line corresponds to the electronic density where the non-interacting nesting condition is satisfied, serving as a guide to the eye to locate the center of the antiferromagnet dome. The origin of the ordered phases becomes transparent by examining the bare susceptibility, which separates into independent intra- and inter-orbital contributions. The intra-orbital contribution is diagonal in the orbital index and develops its strongest peak at $\boldsymbol{Q}=(\pi,\pi)$. This peak occurs when the effective chemical potential, $\mu_{\rm eff}$ (the chemical potential including the Hartree-shift for $\Delta_{\rm cf}=t_{\rm inter}=0$), equals $-\tilde{\Delta}_{\rm cf}/2$ for the lower orbital or $\tilde{\Delta}_{\rm cf}/2$ for the upper orbital. As illustrated in \cref{fig:fs_instabilities_crystal_field}~(a), these conditions correspond to perfect intra-band Fermi-surface nesting. The associated enhancement of the intra-orbital susceptibility drives the Néel instability and explains why the magnetic dome shifts toward lower electron densities with increasing crystal-field splitting, as observed in the phase diagram of \cref{fig:delta_t_inter}(c).

The mechanism underlying the Et phase is qualitatively different. At half-filling and for $\tilde{\Delta}_{\rm cf}\gtrsim t$, the dominant instability occurs in the inter-orbital channel. As shown in \cref{fig:fs_instabilities_crystal_field}~(b), the ordering vector $\boldsymbol{Q}=(\pi,\pi)$ connects occupied states in the lower-energy band with unoccupied states in the upper band, producing strong inter-band nesting. The resulting enhancement of the inter-orbital susceptibility drives coherent particle-hole condensation between the two orbitals, thereby stabilizing the Et phase. 

\begin{figure}
    \centering
    \includegraphics{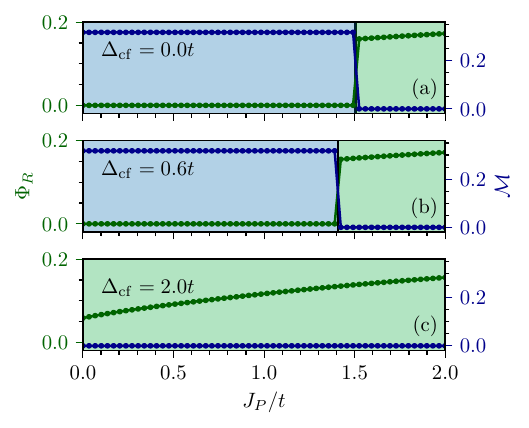}
    \caption{Strength of the Et and intra-orbital magnetic order parameters as a function of the pair-hopping for an increasing strength of the crystal field splitting across panel (a), (b) and (c). The dashed black line signals the transition point between the two different phases. Other parameters: $L=20$, $U=3.0t$, $U^\prime=2.0t$, $J_H=0.5t$, $T=0.1t$ and $n_{e}=2$.}
    \label{fig:J_P_vs_crystal_field_triplet}
\end{figure}

It is natural to ask whether the pair-hopping interaction is strictly necessary to stabilize the Et phase in the presence of the crystal field splitting, which shifts the instability towards intra-orbital magnetic order to lower densities. We address this in \cref{fig:J_P_vs_crystal_field_triplet}, where the strength of the Et order parameter is plotted as a function of $J_P$ for increasing $\Delta_{\rm cf}$. As shown by the discontinuous behavior of the order parameters, increasing either $\Delta_{\rm cf}$ or $J_P$ drives a direct phase transition from an orbital-aligned Néel antiferromagnetic state into the Et phase. Crucially, panel (c) demonstrates that an Et dome robustly develops at half-filling, even in the complete absence of pair-hopping, if the crystal-field splitting is strong enough. To understand why the Et order survives without $J_P$, we contrast this regime with the $\Delta_{\rm cf} = 0$ case discussed in the previous section, where no extended Et order emerges when $J_P = 0$. The mechanism stabilizing the excitonic phase here relies on the crystal-field splitting. A finite $\Delta_{\rm cf}$ shifts the magnetic dome toward regions of lower electronic density, leaving the half-filled regime dominated by inter-orbital instabilities. The non-interacting susceptibility diverges at $T=0$ for the nesting vector $\boldsymbol{Q}=(\pi,\pi)$, this inter-orbital instability persists entirely independent of $J_P$. Physically, the crystal field energetically favors the double occupation of the lower orbitals, which directly competes with the intra-orbital Hubbard repulsion. It is this competition, mediated by the inter-orbital susceptibility divergence at half-filling, that ultimately stabilizes the Et dome without the need for pair-hopping.

We now turn to the effect of an inter-orbital hybridization $t_{\rm inter}$, which acts as an explicit symmetry-breaking term for the relative orbital parity. This hybridization generates an effective inter-orbital exchange that favors a local spin-singlet to optimize kinetic energy, competing directly with the ferromagnetic Hund's coupling that aligns the orbital magnetic moments into a spin-triplet. \cref{fig:delta_t_inter}~(b) displays the resulting phase diagram in the plane of $t_{\rm inter}$ and electron density. At half-filling, increasing $t_{\rm inter}$ drives a transition from an orbitally aligned N{\'e}el phase to an orbitally anti-aligned N{\'e}el state at a critical value $t_{\rm inter}^\star$. These states are distinguished by symmetric and antisymmetric spin combinations of the intra-orbital spins. As $\boldsymbol{S}_{\rm sym}$ and $\boldsymbol{S}_{\rm asym}$ transform under different irreducible representations of the underlying symmetry group (see \cref{tab:order_parameter_irreps}), they constitute distinct thermodynamic phases: the aligned N{\'e}el phase exhibits $\langle \boldsymbol{S}_{\rm sym}\rangle \neq 0$ with $\langle \boldsymbol{S}_{\rm asym}\rangle = 0$, whereas the anti-aligned phase has $\langle \boldsymbol{S}_{\rm sym}\rangle = 0$ and $\langle \boldsymbol{S}_{\rm asym}\rangle \neq 0$.

Although this phase diagram differs qualitatively from the one induced by a crystal-field splitting, the underlying mechanisms can be understood within the same RPA framework. As shown in \cref{fig:delta_t_inter}~(d), two distinct ordered domes emerge as a function of doping. This structure originates from two distinct Fermi-surface nesting conditions realized at and away from half-filling. The electron density of the center of the dome above $(+)$ and below $(-)$ half-filling is given by a similar \cref{eq:dome_density_cf} with the crystal-field term swapped by $\tilde{t}_{\rm inter}$, the inter-orbital hybridization corrected by the Hartree energy shift (see Appendix~\ref{subsec:mean_field_paramagnetic}).

As inter-orbital hybridization splits the non-interacting bands into bonding and anti-bonding states, it strongly mixes orbital character; consequently, the dominant nesting vector at half-filling connects the bonding and anti-bonding Fermi surfaces rather than the original unhybridized orbitals. For the representative value of $t_{\rm inter}$ shown in \cref{fig:delta_t_inter}~(d), the dome centered at half-filling is characterized by finite $\langle \boldsymbol{S}_{\rm asym} \rangle$. In contrast, the dome centered away from half-filling simultaneously develops non-zero $\langle \boldsymbol{S}_{\rm sym} \rangle$ and Et order $\langle \boldsymbol{\Psi} \rangle$. As the reduced symmetry forces $\boldsymbol{S}_{\rm sym}$ and $\boldsymbol{\Psi}_R$ to transform under the same irreducible representation, they label the same phase, and generically emerge together. Away from the centers of these domes, finite doping stabilizes an extended region that exhibits intertwined spin-density and Et density waves.

\section{Conclusion} 
\label{sec:conclusions}

In this work, we presented a comprehensive unrestricted Hartree-Fock study, complemented by an analytical RPA instability analysis, of the two--orbital Hubbard-Kanamori model on a square lattice. Rather than confining our analysis to the rotationally invariant regime characteristic of isolated atomic $d$-shells, we treated the intra- and inter-orbital repulsions, Hund's coupling, and pair-hopping amplitudes as independent parameters. This unconstrained approach allowed us to systematically elucidate their individual roles relative to a representative intermediate-coupling baseline ($U = 3t$, $U' = 2t$, $J_H = J_P = 0.5t$).

By systematically classifying all one-body bilinears according to the irreducible representations of the relevant symmetry group, we established the microscopic symmetry-breaking hierarchy governing intra-orbital magnetic, excitonic, and orbital orders. In the absence of crystal-field splitting and inter-orbital hybridization, the internal orbital degrees of freedom exhibit a $D_4$ symmetry. We demonstrated that the interplay of the Kanamori parameters naturally partitions the ground-state phase diagram into four canonical regimes, whose underlying mechanisms are driven by distinct RPA nesting channels. 

While the intra-orbital magnetic regime largely mirrors single-band Hubbard phenomenology, featuring a commensurate N{\'e}el dome at half-filling that gives way to spiral and collinear spin-charge stripes upon doping, the orbital degree of freedom dramatically enriches this landscape. A dominant inter-orbital repulsion stabilizes a $\mathbb{Z}_2$-broken ODW, suppressing local magnetic moments via the formation of intra-orbital doublons. Alternatively, in the intermediate interaction regime, pair-hopping processes drive the formation of Et condensates (for ferromagnetic Hund's coupling) or Es condensates (for antiferromagnetic Hund's coupling). Crucially, these excitonic states spontaneously break the $\mathbb{Z}_2$ relative orbital parity symmetry of the $D_4$ group, hosting their own rich hierarchy of commensurate, incommensurate, and multi-$\mathbf{Q}$ textures upon doping. Near the phase boundaries separating these four canonical regimes, we identified broad parameter windows characterized by deeply intertwined and coexisting magnetic, excitonic, and charge modulations.

Moving beyond the orbitally degenerate limit, we showed that explicit symmetry-breaking fields, namely, crystal-field splitting and inter-orbital hybridization, qualitatively restructure the non-interacting Fermi surface, thereby redirecting the leading instabilities. A crystal-field splitting detunes the intra-orbital nesting at half-filling, shifting the magnetic dome to finite carrier concentrations while pinning an inter-orbital-nested excitonic dome directly at half-filling. Similarly, inter-orbital hybridization stabilizes an orbitally anti-aligned magnetic state at half-filling and gives rise to an intertwined state of orbitally aligned magnetic order and Et order upon doping. Remarkably, the crystal-field splitting stabilizes Et order even in the absence of pair hopping.

Collectively, our results provide a unified, symmetry-resolved map of the two--orbital Hubbard-Kanamori model's mean-field phase diagram across its extended interaction landscape. Motivated by recent developments in infinite-layer nickelates and related multi-orbital quantum materials, where crystal-field effects and inter-orbital interactions often escape strict atomic symmetry constraints, this broader parameter space establishes a valuable reference frame. A natural direction for future research will be to evaluate the stability of these phases, particularly the exotic excitonic stripes and spirals, beyond the mean-field level, and to investigate their direct implications for transport anomalies and unconventional superconducting pairing.

\begin{figure*}[t]
    \centering
    \includegraphics{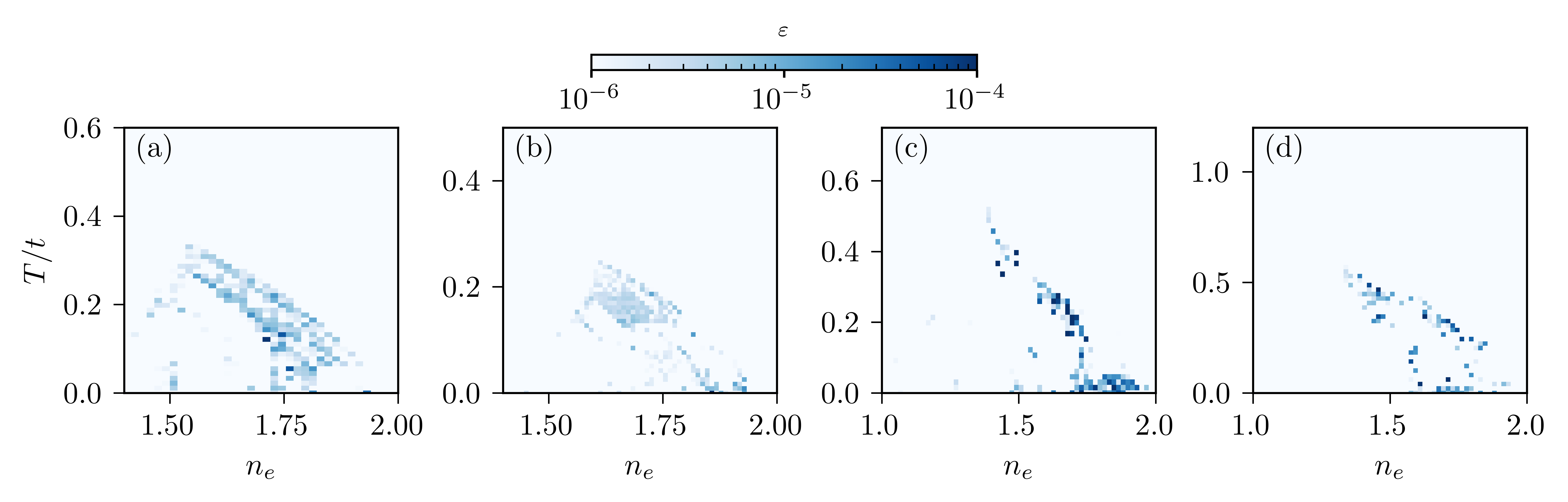}
    \caption{Maximum difference in the local real-space order parameter, $\varepsilon$, in the final mean-field sweep. Panel (a) corresponds to $U=3.0t$, $U^\prime=2.0t$, $J_H=J_P=0.5t$, panel (b) to $U=2.0t$, $U^\prime=2.5t$, $J_H=J_P=0.5t$, panel (c) to $U=2.0t$, $U^\prime=3.5t$, $J_H=J_P=0.5t$ and panel (d) to $U=2.0t$, $U^\prime=3.0t$, $J_H=-0.5t$ and $J_P=t$. Other parameters: $\Delta=0$, $t_{\rm inter}=0$ and $L=20$.}
    \label{fig:error_mean_field}
\end{figure*}

 \begin{acknowledgments}
We thank Shuai Chen, Martin Ulaga, and Andrew Millis for enlightening discussions and comments on the early versions of the draft. A.W. acknowledges support by the German Research Foundation (DFG) through the Emmy Noether program (Grant No. 509755282). Funded by the European Union (ERC, MoNiKa, 101220368). Views and opinions expressed are however those of the author(s) only and do not necessarily reflect those of the European Union or the European Research Council. The numerical calculations presented in this work were performed in the Max Planck Institute for the Physics of Complex Systems computing infrastructure.
\end{acknowledgments}

\appendix

\section{Details of the Mean-Field Calculations}
\label{sec:HF_calculations}

The different mean-field \emph{Ans\"atze} used in this work are based on the Gibbs-Bogoliubov-Feynman inequality~\cite{Feynman_PhysRev.97.660,kuzemsy_variational_principles,Brito_mean_field_2022}. The theorem states that the variational grand potential $\Omega_{V}$, computed with a trial mean-field Hamiltonian $\mathcal{H}_\text{MF}$, provides an upper bound to the exact grand potential $\Omega$ of the corresponding interacting Hamiltonian $\mathcal{H}$:
\begin{equation}
    \Omega \le \Omega_\text{MF} + \left\langle \mathcal{H} - \mathcal{H}_\text{MF} \right\rangle_\text{MF} \equiv \Omega_V,
    \label{eq:variational_principle}
\end{equation}
where $\beta=1/(k_BT)$, $\Omega_\text{MF} = -k_B T \ln \operatorname{Tr} \left\{ e^{-\beta(\mathcal{H}_\text{MF} - \mu \mathcal{N})} \right\}$ is the mean-field grand potential and $\mathcal{N} = \sum_{i,\alpha,\sigma} c^\dagger_{i\alpha\sigma} c_{i\alpha\sigma}$. The  expectation value is computed with respect to the grand-canonical density matrix constructed using the the mean-field state, where for an arbitrary operator, $\mathcal{O}$, we have 
\begin{equation}
\left\langle \mathcal{O} \right\rangle_\text{MF} =\dfrac{\operatorname{Tr} \left\{ \mathcal{O} e^{-\beta(\mathcal{H}_\text{MF} - \mu \mathcal{N})} \right\} }{\operatorname{Tr} \left\{ e^{-\beta(\mathcal{H}_\text{MF} - \mu \mathcal{N})} \right\}}. 
\end{equation}
In our calculations, the chemical potential $\mu$ is determined self-consistently for a given desired target particle number. Therefore, the meaningful quantity to compare across different mean-field states is the variational free energy,
\begin{equation}
    F_{V} = \Omega_V + \mu \langle \mathcal{N} \rangle.
    \label{eq:variational_free_energy}
\end{equation}
We choose $\mathcal{H}_\text{MF}$ to be the most general particle-number-conserving quadratic Hamiltonian in real space:
\begin{equation}
\begin{aligned}
\mathcal{H}_{\rm MF}	&=\mathcal{H}_0 +\sum_{i,j} \sum_{\alpha,\gamma}\sum_{\sigma,\sigma^\prime} \varepsilon^{\rm MF}_{i\alpha \sigma;j\gamma \sigma^\prime} c^\dagger_{i\alpha\sigma} c_{j\gamma\sigma^\prime},
\end{aligned}
\end{equation}
By applying the variational principle of \cref{eq:variational_principle}, we derive the self-consistent equations for the mean-field parameters,
\begin{equation}
   \varepsilon^{\rm MF}_{\boldsymbol{i};\boldsymbol{j}} = \sum_{\boldsymbol{k},\boldsymbol{l}} \left( V^{\boldsymbol{i}\boldsymbol{l}}_{\boldsymbol{k}\boldsymbol{j}} + V^{\boldsymbol{l}\boldsymbol{i}}_{\boldsymbol{j}\boldsymbol{k}} - V^{\boldsymbol{i}\boldsymbol{l}}_{\boldsymbol{j}\boldsymbol{k}} - V^{\boldsymbol{l}\boldsymbol{i}}_{\boldsymbol{k}\boldsymbol{j}} \right) \left\langle c^\dagger_{\boldsymbol{l}} c_{\boldsymbol{k}} \right\rangle_{\rm MF},
   \label{mean_field_selfc_eq}
\end{equation}
where $\boldsymbol{i}\equiv\left(i,\alpha,\sigma \right)$ and $V$ is the iteration tensor, which is local with respect to the lattice site, 
\begin{equation}
    \mathcal{V} = \sum_{j} V^{\alpha_1 \sigma_1; \alpha_2 \sigma_2}_{\alpha_3 \sigma_3; \alpha_4 \sigma_4} c^\dagger_{j,\alpha_1,\sigma_1} c^\dagger_{j,\alpha_2,\sigma_2} c_{j,\alpha_3,\sigma_3} c_{j,\alpha_4,\sigma_4},
\end{equation}
where $\mathcal{V}$ corresponds to the Kanamori Hamiltonian in \cref{eq:kanamori_interaction}.

The self-consistent equations in \cref{mean_field_selfc_eq} were iterated using a linear mixing algorithm, until, either the maximum difference between two successive iterations satisfied $\left|(\varepsilon^{\rm MF}_{\boldsymbol{i},\boldsymbol{j}})^{n+1}-(\varepsilon^{\rm MF}_{\boldsymbol{i},\boldsymbol{j}})^n\right|<10^{-6}$,
or a maximum of $10^4$ iterations was reached. \Cref{fig:error_mean_field} shows the final maximum residual between successive iterations for the four temperature--electron density phase diagrams presented in \cref{fig:T_vs_n_magnetic_dome_w_JP} of the main text. The most challenging region to converge corresponds to the intermediate to low-temperature with a finite-doping. In this regime, the symmetry-broken solutions are characterized by an ordering vector, which may not be commensurate with the finite lattice used in the calculations, thereby hindering convergence. For every point in the phase diagrams, we performed at least ten independent calculations starting from different random initial conditions in an effort to identify the symmetry broken state that had the lowest free-energy.

\section{Further Details on the Internal Symmetries Hubbard-Kanamori Hamiltonian}
\label{appendix:hk_symmetries}
For a generic value of the parameters in the Kanamori interaction, the orbital quantum number transforms under a discrete symmetry group. On the other hand, for specific parameter points, this symmetry group is enriched and possesses continuous symmetries. In this Appendix, we discuss the internal orbital symmetries of the Hubbard-Kanamori model in greater detail and classify under which irreducible representation of the symmetry group the observables defined in \cref{sec:orders_multi_orbital} transform.

First, we show that the symmetry group is the dihedral group of order eight, $D_4$, in the limit $\Delta_{\rm cf}=t_{\rm inter}=0$. To make this explicit, we define the orbital spinor $\psi_{j,\sigma} = (c_{j,1,\sigma}, c_{j,2,\sigma})^T$. Under orbital exchange and relative orbital parity, this spinor transforms, respectively, as
\begin{equation}
    \mathcal{X} \psi_{j,\sigma} \mathcal{X}^{-1} = \tau^x \psi_{j,\sigma},\quad \mathcal{P} \psi_{j,\sigma} \mathcal{P}^{-1} = \tau^z \psi_{j,\sigma},
\end{equation}
where $\tau^x$ and $\tau^z$ act on the orbital subspace. These two operations generate all eight elements of the $D_4$ group: $\{I, \tau^x, \tau^z, -\tau^x, -\tau^z, \tau^x\tau^z, -\tau^x\tau^z, -I\}$. For completeness, in \cref{tab:d4_character_table}, we list the characters of the irreducible representations of $D_4$, which are also used in the main text to classify how the local observables transform under these internal symmetries.

Now, we focus on specific high symmetry points of the two-orbital Hubbard-Kanamori. As mentioned in the main text, in the absence of the pair-hopping and $t_{\rm inter}$, the orbital symmetry group is enriched from $D_4$ to U(1) as the Hamiltonian is invariant under the unitary transformation generated by $\tau^z$ of the fermionic operators,
\begin{equation}
\begin{aligned}
   e^{i\theta/2 \tau_z} c_{j,1,\sigma}e^{-i\theta/2 \tau_z} &=  c_{j,1,\sigma} e^{-i\theta/2},\\ e^{i\theta/2 \tau_z}c_{j,2,\sigma}e^{-i\theta/2 \tau_z} &=  c_{j,2,\sigma} e^{i\theta/2}.
\end{aligned}
\end{equation}
In this case, the putative order parameters transform under the irreducible representation of 
\begin{equation}
G=(\text{U}_{c}(1)\times \text{SU}(2)\times \text{U}_{o}(1)) / \left( \mathbb{Z}_2 \times \mathbb{Z}_2 \right) \rtimes \mathbb{Z}^T_2    
\end{equation}
which are identified in the first column of \cref{tab:order_parameter_irreps_appendix}. All bilinear operators involving fermionic operators acting on the same orbital transform under the charge zero representation, while the complex Et operator and complex Es operator transform under the charge 1 representation. In this case, the real and As such, for both triplet and singlet excitonic order to form, the continuous U(1) orbital symmetry must be spontaneously broken. On the other hand, if $t_{\rm inter}$ is finite and $\Delta_{\rm cf} \neq 0$ all orbital symmetries are explicitly broken. Another relevant limit corresponds in having $\Delta=t_{\rm inter}=J_P=0$. In this case, the full orbital symmetry group is isomorphic to O(2) as it is generated by the continuous U(1) symmetry combined with the discrete orbital permutation symmetry. 

There is another relevant regime where the model acquires an orbital O(2) symmetry. This corresponds to the special case having $\Delta=t_{\rm inter}=0$, $U=U^\prime-2J_H$ and $J_P=J_H$ as in this case the Kanamori interaction is invariant under a local basis rotation of the orbital degrees of freedom with improper transformations as well. The local basis rotation is generated by $\tau^y$,
\begin{equation}
\begin{aligned}
e^{i \theta \tau^y} \psi_{j,\sigma}e^{-i \theta \tau^y} = \begin{pmatrix}
    \cos(\theta) &  \sin(\theta)\\
    -\sin(\theta) &\cos(\theta) \\
\end{pmatrix}\psi_{j,\sigma}.
\end{aligned}
\end{equation}

\begin{table}[t]
    \centering
    \renewcommand{\arraystretch}{1.2}
        \begin{tabular}{|l | c | c | c | c | c|}
         \hline\hline
            $D_4$ & $I$ & $\{ \pm \tau^x\tau^z \}$ & $-I$ & $\{ \pm \tau^x \}$ & $\{ \pm \tau^z \}$ \\
            \hline
            $A_1$ & $1$ & $1$  & $1$  & $1$  & $1$  \\\hline
            $A_2$ & $1$ & $1$  & $1$  & $-1$ & $-1$ \\\hline
            $B_1$ & $1$ & $-1$ & $1$  & $1$  & $-1$ \\\hline
            $B_2$ & $1$ & $-1$ & $1$  & $-1$ & $1$  \\\hline
            $E$   & $2$ & $0$  & $-2$ & $0$  & $0$ \\
            \hline\hline
        \end{tabular}
    \caption{Character table for the dihedral group $D_4$. The five conjugacy classes are expressed explicitly in terms of the orbital Pauli matrices $\tau^a$, which act on the orbital spinor $\psi_{j,\sigma}$.}
    \label{tab:d4_character_table}
\end{table}

\section{Classification of Local Order Parameters in Symmetry-Broken States}
\label{app:cat_orbital_components}

\begin{table}[t]
    \caption{Classification of local bilinear observables by the irreducible representations of the symmetry group of the two-orbital Hubbard-Kanamori in regimes with $J_P=0$ and $t_{\rm inter}=0$. The charge, spin, and time-reversal symmetries form the group $\mathcal{G}=\mathrm{U}_c(1) \times \mathrm{SU}(2) \rtimes \mathbb{Z}^T_2$, which acts independently of the orbital symmetries. All listed observables belong to the charge sector $C=0$ and are grouped by spin representation $S$. Table entries $(O, \mathcal{T})$ denote the orbital representation and time-reversal parity ($\mathcal{T}=\pm$). For the orbital sector $O$, integer charges label $\mathrm{U}_o(1)$ irreps (with complex conjugate pairs denoted by direct sums $1 \oplus -1$), while $A_1, A_2, E_1$ label the $\mathrm{O}(2)$ irreps.}
    \label{tab:order_parameter_irreps_appendix}
    \begin{ruledtabular}
        \begin{tabular}{l c c}
            & \multicolumn{2}{c}{Symmetry Group} \\
            \cline{2-3}
            & $t_{\rm inter} = J_P = 0$ & $\Delta = t_{\rm inter} = J_P = 0$ \\
            \multicolumn{1}{c}{Observables} 
            & $\mathcal{G}\times \mathrm{U}_o(1)$ 
            & $\mathcal{G} \times \mathrm{O}(2)$ \\
            \hline
            $S=0 \quad \tau^z$                        & $(0,+)$ & $(A_2,+)$ \\
            $\phantom{S=0 \quad} \Psi_R$              & $(1 \oplus -1,+)$ & $(E_1,+)$ \\
            $\phantom{S=0 \quad} \Psi_I$              & $(1 \oplus -1,-)$ & $(E_1,-)$ \\
            \hline
            $S=1 \quad \boldsymbol{S}_{\rm sym}$              & $(0,-)$ & $(A_1,-)$ \\
            $\phantom{S=1 \quad} \boldsymbol{S}_{\rm asym}$    & $(0,-)$ & $(A_2,-)$ \\
            $\phantom{S=1 \quad} \boldsymbol{\Phi}_R$ & $(1 \oplus -1,-)$ & $(E_1,-)$ \\
            $\phantom{S=1 \quad} \boldsymbol{\Phi}_I$ & $(1 \oplus -1,+)$ & $(E_1,+)$ \\
        \end{tabular}
    \end{ruledtabular}
\end{table}

\begin{table*}[t]
\centering
\begin{tabular}{ll ccc c p{0.22\textwidth}}
\hline\hline
\textbf{Phase Family} & \textbf{Final Phase Label} & $\mathbf{q}_S$ & $\mathbf{q}_C$ & $\mathbf{q}_E$ & \textbf{OS} & \textbf{Physical Description} \\
\hline

\textbf{Disordered} & \textbf{PM/PO} & $0$ & $0$ & $0$ & Preserved & Uniform paramagnetic and paraorbital state. \\
\hline

& \textbf{Néel} & $(\pi,\pi)$ & $0$ & $0$ & Preserved & Commensurate antiferromagnetic order. \\
\textbf{Magnetic} & \textbf{Spin Spiral} & Spiral $Q$ & $0$ & $0$ & Preserved & Non-collinear spin spiral. \\
& \textbf{S/C Stripe/Multi-$Q$} & $sQ$/$mQ$ & $0$ or $Q$ & $0$ & Preserved & Collinear or multi-$Q$ spin density wave with optional charge modulation. \\
\hline

& \textbf{Et $(\pi,\pi)$} & $0$ & $0$ or $Q$ & $(\pi,\pi)$ & Preserved & Commensurate checkerboard Et order. \\
\textbf{Triplet} & \textbf{Néel+Et $(\pi,\pi)$} & $(\pi,\pi)$ & $0$ or $Q$ & $(\pi,\pi)$ & Preserved & Coexisting Néel spin and commensurate Et order. \\
\textbf{Excitonic} & \textbf{Et Stripe/Multi-$Q$} & $0$ & $0$ or $Q$ & $sQ$/$mQ$ & Preserved & Collinear or multi-$Q$ Et density wave. \\
& \textbf{Et Spiral} & $0$ & $0$ or $Q$ & Spiral $Q$ & Preserved & Non-collinear Et spiral. \\
\hline

\textbf{Coexistence} & \textbf{S/C/Et/Es} & $Q \neq 0$ & $Q \neq 0$ & $Q \neq 0$ & Preserved & Full coexistence of spin, charge, triplet and Es orders. \\
& \textbf{Et/C} & $0$ & $Q \neq 0$ & $Q \neq 0$ & Preserved & Coexisting Et and charge density wave. \\
\hline

\textbf{Orbital} & \textbf{ODW $(\pi,\pi)$} & $0$ & $0$ or $Q$ & $0$ & Broken & Commensurate alternating orbital polarization. \\
\textbf{Ordered} & \textbf{ODW Stripe/Multi-$Q$} & $0$ & $0$ or $Q$ & $0$ & Broken & Incommensurate or multi-$Q$ orbital polarization wave. \\
\textit{(OS Broken)} & \textbf{OS} & Any & Any & Any & Broken & Any complex mixed phase with broken orbital symmetry. \\
\hline

& \textbf{Es $(\pi,\pi)$} & $0$ & $0$ or $Q$ & $(\pi,\pi)$ & Preserved & Commensurate checkerboard Es order. \\
\textbf{Singlet} & \textbf{Es Stripe/Multi-$Q$} & $0$ & $0$ or $Q$ & $sQ$/$mQ$ & Preserved & Collinear or multi-$Q$ Es density wave. \\
\textbf{Excitonic} & & & & & & \\
\hline

\textbf{Unclassified} & \textbf{Unknown} & --- & --- & --- & --- & Phase detected but not yet classified by the algorithm. \\
& \textbf{Unconverged} & --- & --- & --- & --- & Hartree-Fock iteration did not converge; solution unreliable. \\
\hline\hline
\end{tabular}
\caption{Phase classification criteria based on momentum-space signatures of the order parameters: Spin ($\mathbf{q}_S$), Charge ($\mathbf{q}_C$), Excitonic ($\mathbf{q}_E$), and Orbital Symmetry (OS). Here $Q$ denotes a generalized finite wavevector; $sQ$ and $mQ$ denote single-$Q$ and multi-$Q$ configurations respectively. Triplet (Et) and singlet (Es) excitonic orders are distinguished. }
\label{tab:phase_classification}
\end{table*}

As discussed in \cref{sec:orders_multi_orbital}, we classify the different phases by computing the Fourier decomposition of the relevant local order parameters. From that, we classify the phase proceeding in the following way. First, we evaluate whether a given order parameter exhibits a significant magnitude. Using Parseval's theorem, we quantify this by computing the total intensity per site,
\begin{equation}
    \mathcal{I} =\sqrt{\sum_{\boldsymbol{Q}} I_{\boldsymbol{Q}}},\quad I_{\boldsymbol{Q}} = \sum_{\lambda} \left| \mathcal{O}^{\lambda}_{\boldsymbol{Q}} \right|^2.
\end{equation}
This gives the total weight of the order parameter. We compare the intensity of the different orders, identifying the dominant one, $\mathcal{I}_{\rm dominant}$, and assume that secondary orders are present if they satisfy
\begin{equation}
    \dfrac{\mathcal{I}}{\mathcal{I}_{\rm dominant}} > 0.2.
\end{equation}
This discards parasitic orders that do not dominate the physics. Next, we identify the dominant Fourier components of the non-negligible orders to determine whether there is a single dominant ordering vector or a set of multiple dominant vectors. If a single dominant vector exists, we check if it coincides with the $M=(\pi,\pi)$ or the $\Gamma=(0,0)$ point in the First Brillouin Zone. These correspond, respectively, to staggered order along both the $x$ and $y$ directions, and uniform order throughout the lattice. We note that when performing the Fourier transform of the observables, a peak at a single $+Q^\ast$ in the FBZ must be accompanied by a peak at $-Q^\ast$ due to the reality of the order parameter. Therefore, even though there are two peaks, this phase is identified as having a single peak.

In order to identify the spin spiral and the Et spiral phase, we first verify if the amplitude of the order parameter remains approximately constant through the lattice up to a relative maximum deviation of $1\%$. To distinguish non-collinear spiral phases from purely collinear (staggered) phases, we perform a singular value decomposition on the real-space order parameter matrix $\mathbb{O}$. This is a $N \times 3$ matrix, where each row corresponds to the three-component local ordering vector at site $j$. For any purely collinear phase, the local moments at all sites must point along a single, fixed global axis $\hat{\mathbf{n}} = (n_x, n_y, n_z)$. The order parameter matrix can therefore be factored into the outer product of a spatial amplitude vector $\mathbf{c}$ and $\hat{\mathbf{n}}$,
\begin{equation}
    \mathbb{O} = \mathbf{c} \otimes \hat{\mathbf{n}} = 
    \begin{pmatrix}
        c_1 \\
        c_2 \\
        \vdots \\
        c_N
    \end{pmatrix} 
    \begin{pmatrix}
        n_x & n_y & n_z
    \end{pmatrix}.
\end{equation}
As such, $\mathbb{O}$ is strictly a rank-1 matrix, and its SVD only has a single non-zero singular value. In contrast, if the system stabilizes a coplanar order, the local vectors span a two-dimensional plane, resulting in a rank-2 matrix with two finite singular values as
\begin{equation}
    \mathbb{O} = \mathbf{c}_1 \otimes \hat{\mathbf{n}}_1 + \mathbf{c}_2 \otimes \hat{\mathbf{n}}_2
\end{equation}
Finally, for a fully non-coplanar order, the matrix achieves full rank, and all three singular values are non-zero. We evaluate the ratio of the first two leading singular values against a numerical tolerance of $10^{-2}$ to exclude collinear phases from the spiral phase classification.

An extensive description of all the labeled phases can be found in \cref{tab:phase_classification}.

\begin{figure*}[t]
    \centering
    \includegraphics{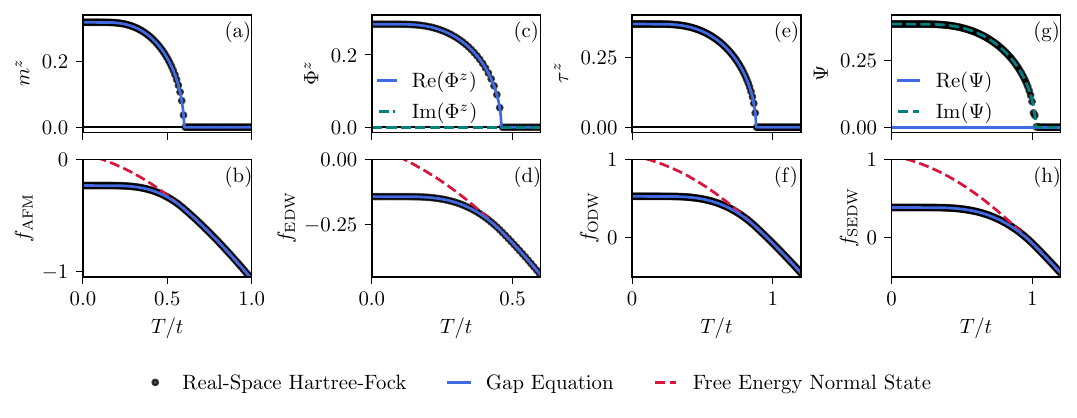}
    \caption{Comparison of the temperature dependence of the mean-field order parameter and variational free energy between the real-space Hartree-Fock calculation and the constrained momentum \emph{anzatz} for the orbital aligned Néel order ((a) and (b)), staggered Et order ((c) and (d)), staggered ODW ((e) and (f)) and staggered Es order ((g) and (h)). The calculations are done with the same parameters used in the main text: $U=3.0t$, $U^\prime=U-2J_H$ and $J_H=J_P=0.5t$ in panel (a), $U=2.0t$, $U^\prime=2.5t$ and $J_H=J_P=0.5t$ (panel (b)), $U=2.0t$, $U^\prime=3.5t$ and  $J_H=J_P=0.5t$ (panel (c)) and $U=2.0t$, $U^\prime=3.0t$ and  $J_H=-0.5t$ and $J_P=t$ in panel (d). Other parameters: $L=10$, $t_{\rm inter}=\Delta_{\rm cf}=0$.}
    \label{fig:benchmarks_HF}
\end{figure*}

\section{Mean-field \emph{Ans\"atze} and RPA Susceptibilities}
\label{appendix:mean_field_anzats_rpa}

Here, we discuss in detail the construction of the mean-field \emph{Ans\"atze} for the N\'eel antiferromagnetic, staggered Et, staggered ODW, and staggered Es phases. These phases provide well-defined benchmarks for validating our real-space Hartree-Fock implementation. Specifically, within their respective regions of stability, we verify that the unrestricted real-space mean-field solution reproduces both the variational free energy and the order parameter obtained from the corresponding constrained \emph{Ansatz}. This is shown in \cref{fig:benchmarks_HF} for the four orders mentioned above.

All four of these phases are characterized by a commensurate ordering wavevector $\boldsymbol{Q}=(\pi,\pi)$, which doubles the spatial unit cell. At half-filling, the mean-field Hamiltonian for each phase corresponds to a two-band model in the magnetic Brillouin zone (MBZ):
\begin{equation}
\mathcal{H}_{\rm MF} = \sum_{\boldsymbol{k} \in \text{MBZ}, \nu} \phi_{\boldsymbol{k},\nu}^{\dagger} 
\begin{pmatrix} 
\varepsilon_{\boldsymbol{k}} & V_{\mathcal{O},\nu} \\ 
V^\ast_{\mathcal{O},\nu} & \varepsilon_{\boldsymbol{k}+\boldsymbol{Q}} 
\end{pmatrix} 
\phi_{\boldsymbol{k},\nu} + E_{\rm bg} + E_{\mathcal{O}}
\end{equation}
where $\varepsilon_{\boldsymbol{k}} = -2t [\cos(k_x) + \cos(k_y)]$ is the non-interacting dispersion, $\nu$ denotes the conserved spin or orbital flavor indices not mixed by the order parameter, and $\Phi_{\boldsymbol{k},\nu}^\dagger$ is the corresponding two-component spinor coupling momentum $\boldsymbol{k}$ to $\boldsymbol{k}+\boldsymbol{Q}$. 

The term $V_{\mathcal{O},\nu}$ represents the off-diagonal mean-field potential responsible for the symmetry breaking. The constant $E_{\rm bg} = -\frac{1}{2}(U + 2U^\prime - J_H)L^2$ at half-filling. $E_{\mathcal{O}}$ is the net Hartree shift associated with the order parameter $\mathcal{O}$.

Diagonalizing this generic Hamiltonian, at half-filling, yields the quasi-particle dispersion $E_{\boldsymbol{k},\nu} = \sqrt{\varepsilon^2_{\boldsymbol{k}} + |V_{\mathcal{O},\nu}|^2}$. The variational free energy per site at half-filling is then strictly given by:
\begin{widetext}
\begin{equation}
f_{\mathcal{O}} = -\frac{T}{L^2} \sum_{\boldsymbol{k} \in \text{MBZ}, \nu, \lambda=\pm} \ln \left(1 + e^{-\beta \lambda E_{\boldsymbol{k},\nu}} \right) + \frac{E_{\mathcal{O}}}{L^2} - \frac{1}{2}\left( U + 2U^\prime - J_H \right)
\end{equation}
\end{widetext}
Minimizing this free energy with respect to the magnitude of the order parameter yields the following gap equation:
\begin{equation}
\frac{1}{L^2}\sum_{\boldsymbol{k} \in \text{MBZ}} \frac{\tanh\left(\beta E_{\boldsymbol{k},\nu} /2\right)}{E_{\boldsymbol{k},\nu}} = \frac{1}{g_{\rm eff}}
\label{eq:gap_equation}
\end{equation}
where $g_{\rm eff}$ is the effective coupling constant responsible for the instability in the respective channel. In the following subsections, we define $V_{\mathcal{O},\nu}$, $E_{\mathcal{O}}$, and $g_{\rm eff}$ for each ordered phase.

\subsection{Antiferromagnet Order}
\label{subsec:antiferromagnet_order}

In the N\'eel antiferromagnet phase, the Hund's coupling aligns the spins within each orbital ferromagnetically, while the moments alternate across sites without net orbital polarization ($\langle T^z_j \rangle = 0$). Assuming spin-SU(2) symmetry is broken along the $z$-axis, the order parameter is the staggered magnetization, $m^z$, and 
\begin{equation}
\begin{aligned}
\phi^\dagger_{\boldsymbol{k},\alpha,\sigma} &= (c^\dagger_{\boldsymbol{k},\alpha,\sigma}, c^\dagger_{\boldsymbol{k}+\boldsymbol{Q},\alpha,\sigma}),\\
V_\sigma &= -\sigma(U+J_H)m^z,\\
E_{\mathcal{O}} &= 2(U+J_H)(m^z)^2 L^2.
\end{aligned}    
\end{equation}
In \cref{eq:gap_equation} we have $g_{\rm eff} = U+J_H$. Notably, the Hund's coupling cooperates with the Hubbard repulsion, enhancing the effective interaction. The critical temperature follows the standard mean-field scaling,
\begin{equation}
T_c \propto \exp\left( - 2\pi \sqrt{\frac{t}{U+J_H}} \right)
\end{equation}

\subsection{Triplet Excitonic Phase}

The staggered Et phase breaks both the continuous spin-SU(2) symmetry, the lattice translation symmetry and the relative orbital parity. Assuming a mean-field state polarized along the $z$ direction, the order parameter is 
\begin{equation}
\Phi^z_j = \Phi_0 e^{i\boldsymbol{Q}\cdot \boldsymbol{r}_j}. 
\end{equation}
and
\begin{equation}
\begin{aligned}
\Phi^\dagger_{\boldsymbol{k},\sigma} &= (c^\dagger_{\boldsymbol{k},1,\sigma}, c^\dagger_{\boldsymbol{k}+\boldsymbol{Q},2,\sigma}),\\
V_\sigma &= -\sigma(J_P \Phi^z + U^\prime (\Phi^z)^\ast),\\
E_{\mathcal{O}} &= 2U^\prime|\Phi^z|^2L^2 + J_P([\Phi^z]^2 + [\Phi^z]^{\ast 2})L^2
\end{aligned}   
\end{equation}
Remarkably, the Hund's coupling $J_H$ is absent since its contributions from the density-like and spin-flip exchange channels exactly cancel. As $\Phi^z$ is complex, minimizing the free energy yields two separate gap equations for its real and imaginary components, governed by different effective couplings: $g_{\rm eff}^{\rm Re} = U^\prime + J_P$ and $g_{\rm eff}^{\rm Im} = U^\prime - J_P$. Assuming $J_P > 0$, the real channel reaches the instability threshold first, ensuring the order parameter condenses purely in the real component ($\text{Im}(\Phi^z) = 0$), preserving the orbital permutation symmetry. The critical temperature is given by
\begin{equation}
T_c \propto \exp\left( - 2\pi \sqrt{\frac{t}{U^\prime+J_P}} \right).
\end{equation}

\subsection{Orbital Density Wave}
\label{appendx:mf_anzats_odw}
The ODW phase preserves the spin-SU(2) symmetry, but breaks the local $\mathbb{Z}_2$ orbital permutation symmetry. The order parameter is the staggered orbital occupation imbalance $T^z$ and 
\begin{equation}
\begin{aligned}
\phi^\dagger_{\boldsymbol{k},\alpha,\sigma} &= (c^\dagger_{\boldsymbol{k},\alpha,\sigma}, c^\dagger_{\boldsymbol{k}+\boldsymbol{Q},\alpha,\sigma}),\\
V_\alpha &= -\alpha(2U^\prime - U - J_H)\tau^z,\\
E_{\mathcal{O}} &= 2(2U^\prime - U - J_H)(\tau^z)^2L^2.
\end{aligned}
\end{equation}
In this case, the effective coupling constant is $g_{\rm eff} = 2U^\prime - U - J_H$. In contrast to the antiferromagnet, the critical temperature for the ODW is suppressed by the Hund's coupling
\begin{equation}
T_c \propto \exp\left( -2\pi \sqrt{\frac{t}{2U^\prime - U - J_H}} \right).
\end{equation}

\subsection{Singlet Excitonic Phase}
\label{appendix:mean_fieldSEDW}

The staggered Es phase is characterized by a spin-singlet excitonic order parameter $\Psi$. Similar to the Et phase,
\begin{equation}
\begin{aligned}
\phi^\dagger_{\boldsymbol{k},\sigma} &= (c^\dagger_{\boldsymbol{k},1,\sigma}, c^\dagger_{\boldsymbol{k}+\boldsymbol{Q},2,\sigma}),\\
V &= (2J_H - U^\prime)\Psi^\ast + J_P \Psi,\\
E_{\mathcal{O}} &= 2(U^\prime - 2J_H)|\Psi|^2L^2 - J_P(\Psi^2 + (\Psi^\ast)^2)L^2.
\end{aligned}
\end{equation}
Applying the variational principle yields two independent gap equations for the real and imaginary components,
\begin{equation}
\begin{aligned}
\text{Re}(\Psi) \left[ \frac{1}{L^2} \sum_{\boldsymbol{k} \in \text{MBZ}} \frac{\tanh(\beta E_{\boldsymbol{k}}/2)}{E_{\boldsymbol{k}}} - \frac{1}{U^\prime - 2J_H - J_P} \right] &= 0,\\
\text{Im}(\Psi) \left[ \frac{1}{L^2} \sum_{\boldsymbol{k} \in \text{MBZ}} \frac{\tanh(\beta E_{\boldsymbol{k}}/2)}{E_{\boldsymbol{k}}} - \frac{1}{U^\prime - 2J_H + J_P} \right] &= 0.
\end{aligned}
\end{equation}
Unlike the Et phase, for $J_P > 0$ the strongest effective coupling occurs in the imaginary channel ($g_{\rm eff}^{\rm Im} = U^\prime - 2J_H + J_P$). Consequently, the Es order parameter acquires a purely imaginary expectation value, breaking both time-reversal and orbital permutation symmetries. The critical temperature asymptotically scales with
\begin{equation}
T_c \propto \exp\left( -2\pi \sqrt{\frac{t}{U^\prime - 2J_H + J_P}} \right).
\end{equation}

\subsection{RPA in the High-Temperature State}
\label{subsec:mean_field_paramagnetic}

Finally, we discuss the mean-field \emph{Ansatz} for the high-temperature state. In this regime, the mean-field state is homogeneous and preserves the original symmetries of the Hamiltonian,
\begin{equation}
    \mathcal{H}^{0}_{\rm MF} = \sum_{\boldsymbol{k}\in {\rm FBZ},\sigma}\phi_{\boldsymbol{k}\sigma}^{\dagger}\begin{pmatrix}\varepsilon_{\boldsymbol{k}} - \mu + \dfrac{\tilde{\Delta}_{\rm cf}}{2} & -\tilde{t}_{{\rm inter}}\\
-\tilde{t}^\ast_{{\rm inter}} & \varepsilon_{\boldsymbol{k}} - \mu - \dfrac{\tilde{\Delta}_{\rm cf}}{2}
\end{pmatrix}\phi_{\boldsymbol{k}\sigma},
\label{eq:mean_field_Hamiltonian_paramagnetic}
\end{equation}
where $\phi_{\boldsymbol{k},\sigma} = \begin{pmatrix}
     c_{\boldsymbol{k},1,\sigma} & c_{\boldsymbol{k},2,\sigma}
\end{pmatrix}^T$ and $\tilde{\Delta}_{\rm cf}$ and $\tilde{t}_{\rm inter}$ are the effective crystal field splitting and inter-orbital hopping. In the presence of crystal field-splitting, the orbital permutation symmetry is explicitly broken by the Hamiltonian. This generates an Hartree correction to the bare-bands,
\begin{equation}
\begin{aligned}
 \tilde{\Delta}_{\rm cf} &=\Delta_{\rm cf} + \dfrac{2}{L^2} \sum_{k} \left(2U^\prime-J_H - U\right)T^z_0,\\
 T^z_0 &= \dfrac{1}{2 L^2} \sum_{k}\left( \langle n_{\boldsymbol{k},2,\sigma} \rangle -  \langle n_{\boldsymbol{k},1,\sigma} \rangle \right).
\end{aligned}
\end{equation}
If the bare inter-orbital hopping $t_{\rm inter}$ is zero, the orbital flavor is a good quantum number, and the inter-orbital coherence strictly vanishes. However, a finite $t_{\rm inter}$ explicitly breaks this conservation, hybridizing the orbitals and inducing a finite expectation value $\langle c^\dagger_{j,1,\sigma} c_{j,2,\sigma} \rangle$. Through the Hartree-Fock decoupling of the Kanamori interactions, this bare coherence generates a static self-energy correction that shifts the bare hopping into an effective, interaction dependent parameter
\begin{equation}
\begin{aligned}
\tilde{t}_{\rm inter} &= t_{\rm inter} + \left(U^\prime-2J_H \right) \dfrac{1}{2 L^2} \sum_{\boldsymbol{k},\sigma}\langle c^\dagger_{\boldsymbol{k},2,\sigma} c_{\boldsymbol{k},1,\sigma} \rangle_{\rm MF}+\\ 
&- J_P \dfrac{1}{2 L^2} \sum_{\boldsymbol{k},\sigma}\langle c^\dagger_{\boldsymbol{k},1,\sigma} c_{\boldsymbol{k},2,\sigma} \rangle_{\rm MF}.
\end{aligned}
\end{equation}

Having established the mean-field Hamiltonian for the high temperature paramagnetic and paraorbital phase, we now analyze its stability to determine the critical temperatures that mark the onset of symmetry-breaking order. This is achieved by locating the divergences of the spin and charge susceptibilities, which we compute within the RPA. The application of RPA to multi-orbital Hubbard models is well established and we refer the reader to Refs.~\cite{PhysRevB.75.224509,Graser_2009} for the detailed derivations. We first define the Fourier transformed generalized spin and charge density operators in orbital space as
\begin{equation}
\begin{aligned}
S^z_{\boldsymbol{q},a b} &= \dfrac{1}{2}\sum_{\boldsymbol{k},\mu,\nu} c^\dagger_{\boldsymbol{k},a,\mu}\sigma^z_{\mu\nu}c_{\boldsymbol{k}+\boldsymbol{q},b,\nu},\\
\rho_{\boldsymbol{q},ab} &= \dfrac{1}{2}\sum_{\boldsymbol{k},\sigma} c^\dagger_{\boldsymbol{k},a,\sigma}c_{\boldsymbol{k}+\boldsymbol{q},b,\sigma}.
\end{aligned}
\end{equation}
The corresponding spin and charge susceptibilities are rank-four tensors, given by
\begin{equation}
\begin{aligned}
\left[\boldsymbol{\chi}^{zz} (\boldsymbol{q},\omega) \right]^{a b}_{cd} &= \langle\langle S^z_{\boldsymbol{q},a b} ; S^z_{-\boldsymbol{q}, cd} \rangle\rangle_\omega,\\
\left[\boldsymbol{\chi}^{\rho\rho} (\boldsymbol{q},\omega) \right]^{ab}_{cd} &= \langle\langle \rho_{\boldsymbol{q}, ab} ; \rho_{-\boldsymbol{q}, cd} \rangle\rangle_\omega,
\end{aligned}
\label{eq:bare_susceptibilites}
\end{equation}
where we have introduced the Zubarev double-bracket notation for the retarded Green's function, $\langle\langle \hat{A} ; \hat{B} \rangle\rangle_\omega = -i \int_{0}^{\infty} \langle [\hat{A}(t), \hat{B}(0)] \rangle e^{i\omega t} dt$. Following standard procedures, we re-sum the bubble diagrams to obtain the RPA expressions given in \cref{eq:rpa_equations} of the main text. The spin and charge interaction tensors $\hat{U}_S$ and $\hat{U}_C$ are given by
\begin{equation}
\begin{aligned}
  \hat{U}_S &= \begin{pmatrix}
      U & J_H & 0 &0\\
      J_H & U & 0 &0\\
      0 & 0 & U^\prime &J_P\\
      0 & 0 & J_P &U^\prime
  \end{pmatrix},\\
  \hat{U}_C &= \begin{pmatrix}
      U & 2U^\prime -J_H & 0 &0\\
      2U^\prime -J_H & U & 0 &0\\
      0 & 0 & 2J_H-U^\prime &J_P\\
      0 & 0 & J_P & 2J_H-U^\prime
  \end{pmatrix},
\end{aligned}
\end{equation}
where these matrices are written in the basis, $\left\{ \left(1,1 \right), \left(2, 2 \right), \left(1, 2 \right), \left(2, 1 \right) \right\}$.

The spin and charge susceptibilities of \cref{eq:bare_susceptibilites} in the high-temperature paramagnetic state are computed using the Matsubara Green functions~\cite{mahan2000many,bruus2004many} of the mean-field Hamiltonian in \cref{eq:mean_field_Hamiltonian_paramagnetic}:
\begin{equation}
\begin{aligned}
    \left[\boldsymbol{\chi}_0 \right]^{a b}_{c d}(\boldsymbol{Q}, i\omega_n)
    &= -\frac{T}{L^2}\sum_{\boldsymbol{k},i\nu_n} \mathcal{G}_{0,a,c}\left(\boldsymbol{k},i\nu_n \right) \\
    &\quad \times \mathcal{G}_{0,d,b} \left(\boldsymbol{k} + \boldsymbol{Q}, i\nu_n + i\omega_n \right), 
\end{aligned}
\end{equation}
where $i\omega_n$ is a bosonic Matsubara frequency, and $i\nu_n$ is a fermionic one. The matrix elements $\mathcal{G}_{0,\alpha,\beta}\left(\boldsymbol{k},i\nu_n \right)$ correspond to the components of the bare Matsubara Green's function matrix, which is given by 

\begin{widetext}
\begin{equation}
\hat{\mathcal{G}}_{0}\left(\boldsymbol{k},i\nu_n \right) = \frac{1}{E_{\boldsymbol{k},+}-E_{\boldsymbol{k},-}} \left[ \frac{1}{i\nu_n - E_{\boldsymbol{k},+}}
    \begin{pmatrix}
        E_{\boldsymbol{k},+}-\xi_{\boldsymbol{k},b} & \tilde{t}_{{\rm inter}}\\
        \tilde{t}_{{\rm inter}}^{\ast} & E_{\boldsymbol{k},+}-\xi_{\boldsymbol{k},a}
    \end{pmatrix} 
    - \frac{1}{i\nu_n - E_{\boldsymbol{k},-}}
    \begin{pmatrix}
        E_{\boldsymbol{k},-}-\xi_{\boldsymbol{k},b} & \tilde{t}_{{\rm inter}}\\
        \tilde{t}_{{\rm inter}}^{\ast} & E_{\boldsymbol{k},-}-\xi_{\boldsymbol{k},a}
    \end{pmatrix} \right],
\end{equation}
\end{widetext}
with 
\begin{equation}
    E_{\boldsymbol{k},\pm} = \frac{1}{2}(\xi_{\boldsymbol{k},a}+\xi_{\boldsymbol{k},b}) \pm \frac{1}{2}\sqrt{ (\xi_{\boldsymbol{k},a}-\xi_{\boldsymbol{k},b})^2 + \tilde{t}^2_{\rm inter} }.
\end{equation}
We note that in the absence of $t_{\rm inter}$, the Green function is diagonal in the orbital quantum number and therefore $\left[\boldsymbol{\chi}_0 \right]^{a b}_{c d}=\delta^{a}_{c} \delta^{b}_{d}\left[\boldsymbol{\chi}_0 \right]^{a b}_{a b}$. Moreover, in the absence of also crystal field splitting, the elements of the susceptibility tensor are all equal to the same value, $\left[\boldsymbol{\chi}_0 \right]^{a b}_{c d}=\chi_0 \;\delta^{a}_{c} \delta^{b}_{d}$. $\chi_0$ is given by the Linhard function
\begin{equation}
    \chi_0 (\boldsymbol{Q}, \omega)= \dfrac{1}{L^2}\sum_{\boldsymbol{k}} \dfrac{f_{\rm FD}\left(\xi_{\boldsymbol{k}+\boldsymbol{Q}}\right) - f_{\rm FD}\left(\xi_{\boldsymbol{k}}\right)}{\omega - \left(\xi_{\boldsymbol{k}+\boldsymbol{Q}} -\xi_{\boldsymbol{k}} \right)}.
\end{equation}

\bibliography{bib}

\end{document}